\documentclass[12pt,showpacs,amsmath,amssymb,pre,endfloats]{revtex4}
\usepackage{epsfig} 
\usepackage{ulem}   

\usepackage{calc,ifthen} 
\newcounter{hours}
\newcounter{minutes}
\newcommand{\determinetime}{
\ifthenelse{\thehours > 11}{\newcommand{\meridian}{PM}}{\newcommand{\meridian}{AM}}
\ifthenelse{\thehours > 12}{\setcounter{hours}{\thehours-12}}{}
}
\newcommand{\printtime}{
\ifthenelse{\theminutes < 10}{\thehours:0\theminutes\meridian}{\thehours:\theminutes\meridian}
}

\usepackage{bm} 

\newcommand{\noi}{\noindent}
\newcommand{\dd}[2]{ \frac{\partial {#1}}{\partial {#2}}}

\newcommand{\q}[1]{(\ref{#1})}
\newcommand{\fig}[1]{Figure~\ref{#1}}
\newcommand{\tabl}[1]{Table~\ref{#1}}
\renewcommand{\d}[1]{\mathsf{d}{#1}}
\newcommand{\ave}[1]{\left\langle{#1}\right\rangle}

\newcommand{\be}{\begin{equation}}
\newcommand{\ee}{\end{equation}}
\newcommand{\bea}{\begin{eqnarray}}
\newcommand{\eea}{\end{eqnarray}}

\newcommand{\zd}{\delta}
\newcommand{\ze}{\epsilon}
\newcommand{\zg}{\gamma}
\newcommand{\zl}{\lambda}

\newcommand{\zvt}{\vartheta}

\newcommand{\zR}{I\hskip-3.8pt R}
\newcommand{\zT}{I\hskip-5.6pt T}

\newcommand{\zW}{\Omega}

\newcommand{\zF}{\Phi}

\newcommand {\bdm} {\begin{displaymath}}
\newcommand {\edm} {\end{displaymath}}
\newcommand {\ba}  {\begin{array}}
\newcommand {\ea}  {\end{array}}

\begin{document}
\preprint{}
\title{Transport complexity in simple porous media}
\author{Owen G. Jepps}
\email{jepps@calvino.polito.it}
\affiliation{Lagrange Fellow, Dipartimento di Matematica, Politecnico di Torino, Corso Duca
degli Abruzzi 24, 10129 Torino, Italy}
\author{Lamberto Rondoni}
\email{lamberto.rondoni@polito.it}
\affiliation{Dipartimento di Matematica and INFM, Politecnico di Torino, Corso
Duca degli Abruzzi 24, 10129 Torino, Italy}
\date{\today}


\begin{abstract}
We examine the transport behaviour of non-interacting particles in a simple channel billiard, at equilibrium and
in the presence of an external field. The channel walls are constructed from straight line-segments.
We observe a sensitive dependence on the model parameters of the transport properties, 
which range from sub-diffusive to super-diffusive regimes. In non-equilibrium, we find a transition in the transport 
behaviour between seemingly-chaotic and (quasi-) periodic behaviour. Our results support the view that normal
transport laws do not need chaos, or quenched disorder, to be realized.
Furthermore, they motivate some new definitions of complexity, which are relevant for transport phenomena.

\end{abstract}


\maketitle


\section{Introduction}
\label{introduction}

One of the fundamental aims of statistical mechanics is to shed light upon the relationship between the macroscopic properties of a
system and its underlying microscopic behaviour. In formalising such a relationship, conditions of ``molecular chaos'' are crucial,
and are commonly assumed in order to extract physical properties from microscopic models of macroscopic systems (including from
molecular dynamics simulations). In fact, various ``Chaotic Hypotheses'' (CH) are  either explicitly or implicitly assumed in the
existing literature; for instance Refs.\cite{GalCoh95,PG98} consider the microscopic dynamics as chaotic, in the sense of uniform
hyperbolicity, or of positive Kolmogorov-Sinai entropy, and the CH of Gallavotti and Cohen explicitly states:

\vskip 5pt
\noindent
{\bf CH (Gallavotti-Cohen 1995). }{\it A reversible many-particle system in a stationary state can
be regarded as a transitive Anosov system for the purpose of computing the macroscopic
properties of the system.
}

\vskip 5pt \noi
This hypothesis does not mean that particle systems actually are of the Anosov type, as the
mathematical definition implies \cite{DR89}, because the Anosov property is invoked only to 
obtain results on a limited number of quantities: those 
which characterize the macroscopic state of the system. Therefore, some of the consequences of the
CH are expected to apply to systems which are not Anosov, the Gallavotti Cohen Fluctuation Theorem
(GCFT) being one of these consequences \cite{GalCoh95}.

The Ergodic Hypothesis (EH) plays a similar role in statistical mechanics. The assumption of ergodicity in equilibrium 
systems
allows one to make a formal association between the thermodynamics of a molecular system and its underlying mechanical properties,
despite the fact that ergodicity is notoriously difficult to prove, and in many cases of physical interest it is known
to be violated. Therefore, ergodicity is assumed for the purpose of extracting macrosopic data from microscopic 
models, largely on the basis that thermodynamic data thus generated appear consistent with observations of real 
physical systems.

At present there is a wide body of results on the applicability of the EH, and theoretical explanations
of why it works have been available for quite some time \cite{Kinchin,GG-SMbook}. Differently, the chaotic
hypotheses, recently introduced to describe nonequilibrium systems, are not well understood yet, and the
question of how much microscopic chaos, and which notion of chaos, is sufficient to explain the observed
macroscopic behaviouir remains elusive \cite{ESR05}. For instance, recent results indicate that 
{\em thermodynamic-like}
connections between the microscopic and macroscopic nature of a particle system can be forged for 
weaker-than-chaotic dynamical systems. In particular, papers such as Refs.\cite{LCW02,physd187_0184} show that 
behaviours which look like normal diffusion and heat conduction can be observed in polygonal billiards, 
although their dynamics are not chaotic. Indeed, their topological entropy vanishes, which means that nearby
orbits do not separate at an exponential rate. Nevertheless, pairs of orbits eventually separate, with the 
only exception being pairs of parallel periodic orbits \cite{GKT95}. Consequently, these systems exhibit a certain sensitive dependence
on the initial conditions, and their dynamics may appear highly disordered, indistinguishable to the eye from chaotic motions.

Similarly, diffusive behaviour is observed in non-chaotic one dimensional lattices of maps, and in the Ehrenfest
wind-tree model \footnote{This is one kind of polygonal billiard, with one point particle which moves in the two
dimensional plane, and undergoes elastic collisons with scatterers which have flat sides. Such collisions do not
defocus neighbouring trajectories, hence chaos is absent.}, with spatial quenched disorder \cite{CCFV03,DCVB99}.
Furthermore, the GCFT has been verified for an initial diffusive phase, during transport observed in a nonequilibrium 
version of the periodic Ehrenfest wind-tree model \cite{jsp099_0857}, and in a model of a mechanical ``pump'', with
flat sides \cite{BRMPEJ}.

It is, therefore, of interest to explore such non-defocussing systems in greater detail, in order to
understand more deeply the nature of transport in non-chaotic particle systems, and to determine its
dependence on the geometry of the medium in which transport takes place. In addition to these more
fundamental issues, there is also the practical consideration of the study of transport in porous
media, as technological developments lead to the production of transport membranes designed ever
more precisely on the atomic scale. In systems where the molecular size is of the same order as the
pore width, solid-fluid interactions can dominate intermolecular interactions at laboratory
temperatures and pressures, to the extent that the contribution from these intermolecular
interactions to the overall transport rate can be neglected, and it is sufficient to consider only
the solid-fluid interaction \cite{owen_prl}. In this respect, the systems studied in this paper
resemble transport in microporous membranes at low (but practically relevant) densities, where
interactions are rare. However, such porous media lie at the fringes of our understanding, in terms
of connecting their microscopic dynamics and their macroscopic properties, so that their 
applications can be further developed from an improved understanding of these theoretical issues.
 
In this paper, we consider the mass transport of point particles in a simple two-dimensional
polygonal channel, described in section \ref{system}, consisting of a unit cell that is replicated
infinitely in one dimension, and bounded in the other dimension. Despite being arguably the simplest
particle system that could be conceived, we observe (in section \ref{results}) a surprising array of
transport behaviours --- while certain properties of the system demonstrate almost trivial
behaviour, other properties display an unpredictable richness, which can be expressed through a
sensitive dependence of the macroscopic behaviour on the parameters which define the geometry of the
boundary. 

In our view, the dependence of the mass transport on the geometry of the system characterizes such 
transport as {\it ``complex''}. Notions of complexity of billiards dynamics, based on symbolic dynamics,
are commonly considered, cf.\ \cite{ST98,GT04}, but in the context of the present paper, it seems 
more appropriate to focus on the complexity of the mass transport as a function of the parameters 
of the system, rather than the complexity of the dynamics for one given set of parameters.
Concerning this aspect, there are various studies which show that noninteracting particle systems, or
lattices of maps, have highly irregular dependences of their transport coefficients on the values of
the paramenters which define them (see, for instance, \cite{LlNiRoMo95,HKG02,ZK04,Kla01,Kla10}). These studies
concern chaotic dynamical systems, and show that the diffusion coefficient, or the conductivity, can be
quite irregular functions of the applied field, of the shape, or of the map's slope. However, in systems 
of particles where these irregularities has been observed, their effects do not preclude important transport 
properties that are well-known in more regular systems. For instance, the irregular behaviour does not 
necessarily prevent a linear regime, close to equilibrium 
\cite{POSITI}. 

In non-chaotic systems of particles, like the ones considered here, the level of irregularity can grow one 
step further; it can produce discontinuous and, in fact, even {\em singular} transport coefficients as
functions of the geometry of the systems. If this is the case, not only is the behaviour of the single 
trajectories unpredictable, and the dynamics is ``complex'' \cite{BCFV02}, but the overall transport
itself becomes largely unpredictable. In other words, in the case of chaotic dynamics,
the unpredictability of the fate of a single initial condition is not reflected in the behaviour of 
an ensemble of initial conditions, while here the behaviour of the ensemble itself --- i.e.\ the object of
interest in transport applications --- is unpredictable. In fact, slight differences in the manufacturing 
of porous membranes may result in totally different transport properties. This leads us to the 
notion of {\em ``transport complexity''}, which will be discussed in section \ref{complexity}.

The dependence of the mass transport on the geometry, i.e.\ of the statistical properties of 
the dynamics on the parameters defining the system, must be connected with the structure of
the ergodic and mixing properties of these systems. The ergodic properties are only known for
systems corresponding to various subsets of the parameter space --- however, we have no
constructive way of explicitly specifying
\emph{which} parameter subsets (cf. Ref.\cite{EG03}). Moreover, the ergodicity alone
does not distinguish very finely between different
transport behaviours. There are various numerical works on the ergodic, mixing and transport
properties of irrational polygons, see, for instance,
Refs.\cite{ACG97,LCW02,CP99,AC03,physrep290-037,physd187_0184,EG86}. However, the models are limited in
number, the triangular billiards being the most studied models, and many questions concerning them
remain open.  Our results can be summarized as follows:
\begin{itemize}
\item We find that diffusive behaviour is straightforward to obtain, despite the
absence of chaos (in the sense of positive Lyapunov exponents), and the absence 
of quenched disorder. 
\item In the nonequilibrium case we find that non-chaotic dynamics makes
the use of Gaussian thermostats physically inappropriate. Indeed, the existence
of a linear regime is not guaranteed in such systems.
\item We find a strong dependence of the transport laws on the geometry of the
system. This shows that these non-chaotic models of noninteracting particles 
cannot be considered as proper thermodynamic systems.
\item We propose three ways of quantifying the dependence of the transport
laws on the geometry of the systems, i.e.\ their {\em transport complexity}.
\end{itemize}


\section{System Details}
\label{system}

Let us begin by recalling the notion of a polygonal billiard.

\vskip 5pt \noi
{\bf Definition 1. }{\it Let ${\mathcal P}$ be a bounded domain in the Euclidean plane $\zR^2$ or on
the standard torus $\zT^2$, whose boundary $\partial {\mathcal P}$ consists of a finite number of 
(straight) line segments. A {\em polygonal billiard} is a dynamical system generated by
the motion of a point particle with constant unit speed inside ${\mathcal P}$, and with elastic 
reflections at the boundary $\partial {\mathcal P}$.
}

\vskip 5pt \noi 
As usual, elastic reflection means that the angle between the incoming velocity and the normal 
to $\partial {\mathcal P}$, at the collision point, equals the angle between the outgoing velocity 
and the same normal. In the general theory of polygonal billiards, ${\mathcal P}$ is not required to 
be convex or simply connected; the boundary may contain internal walls. If the trajectory hits a 
corner of ${\mathcal P}$, in general it does not have a unique continuation, and thus it normally stops
there. 

\vskip 5pt \noi
In continuous time, the dynamics are represented by a flow $\{S^t\}_{t \in (-\infty,\infty)}$ in the phase
space ${\mathcal M} = {\mathcal P} \times \zT^1$, where $\zT^1$ is the unit circle of the velocity angles $\zvt$. 
Because the dynamics are Hamiltonian, the flow preserves the standard measure $dx dy d\zvt$ (with $x,y\in\zR$ or $\zT^2$, such that $(x,y)\in\mathcal{P}$). In discrete time, the dynamics are represented by the ``billiard map'' $\phi$, on the phase space $\Phi = \{ (q,v) \in {\mathcal M} : q \in \partial {\mathcal M}, \langle v , n(q) \rangle \ge 0 \}$ where $n(q)$ is the inward normal vector to $\partial {\mathcal M}$ at $q$, and $\langle \cdot , \cdot \rangle$ is the scalar product. Therefore $\phi$ is the first return map, and the $\phi$-invariant measure on $\Phi$ induced by $dx dy d\zvt$ is $\sin \zvt d\zvt ds$, if $s$ is the arclength on $\partial {\mathcal P}$. Unfortunately, these standard measures are not necessarily selected by the dynamics, in the sense that different absolutely continuous measures are not necessarily evolved towards them. This weakens considerably the importance of the standard measures in the case of non-ergodic polygons.

\vskip 5pt
\noi
{\bf Definition 2. }{\it A polygon ${\mathcal P}$ is called {\em rational} if the angles between 
its sides are of the form $\pi m/n$, where $m,n$ are integers. It is called {\em irrational} otherwise.}

\vskip 5pt
\noi
While polygonal billiards are easily described, their dynamics are extremely difficult to
characterise. For instance it is known that rational polygons are not ergodic, and that they possess
periodic orbits. But it is not known whether generic irrational polygons have any periodic orbit. On the
other hand it is known that irrational polygons whose angles admit a certain superexponentially fast
rational approximation are ergodic \cite{EG03}. In particular the ergodic polygonal billiards are a
dense $G_\zd$ set \footnote{A subset $Y$ of a compact metric space $X$ is a dense $G_\zd$ set of $X$,
if $Y$ is a countable intersection of dense open subsets of $X$.} in any compact set ${\mathcal
Q}$ of polygons with a fixed number of sides, such that the rational polygons with angles with arbitrarily
large denominators are dense in ${\mathcal Q}$ \cite{EG96}. It is not the purpose of this paper to
review exhaustively the properties of polygonal billiards, therefore we refer to the cited
literature for further details.

\begin{figure}
  \epsfig{file=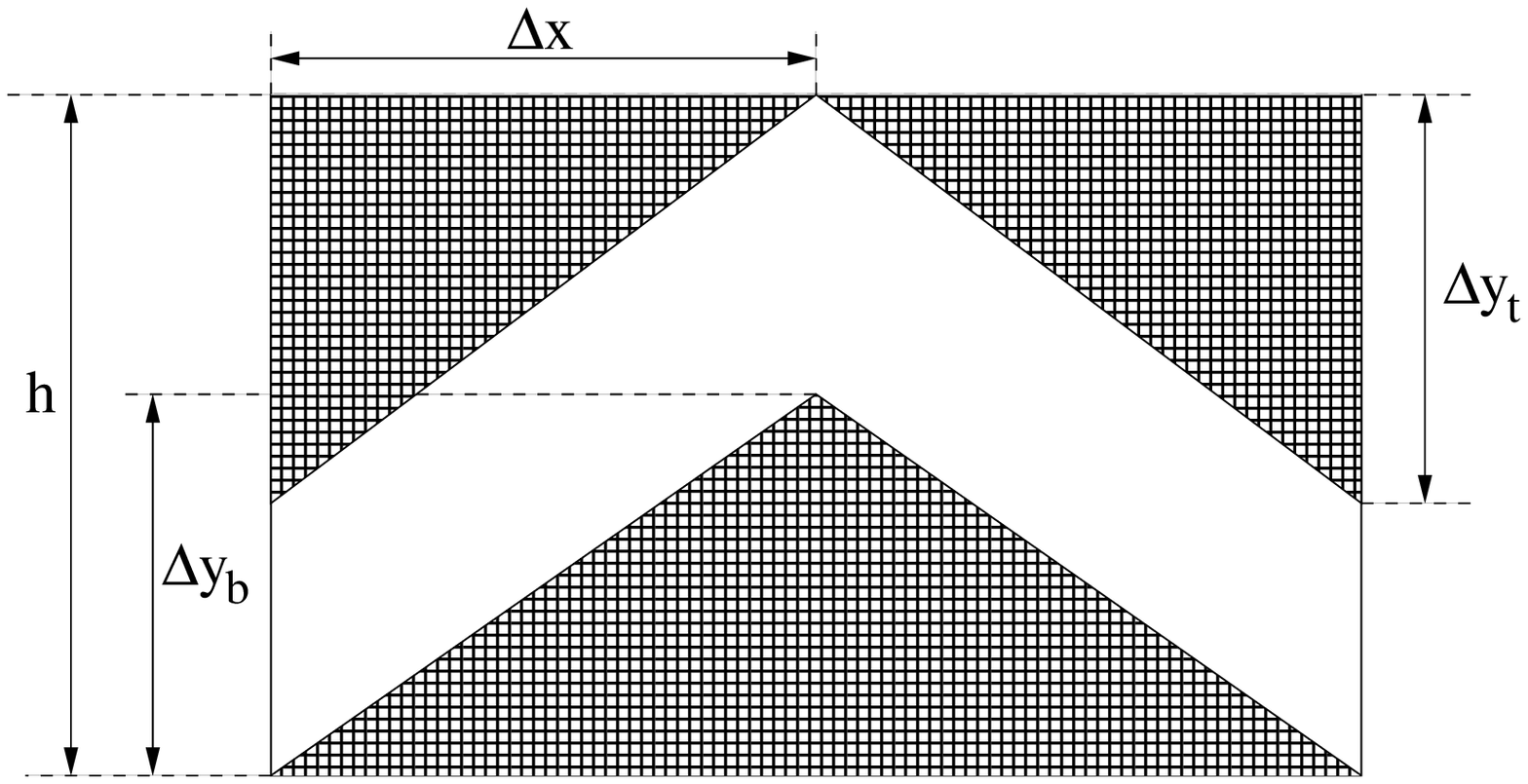 , width=9cm }
  \caption{Symmetric saw-tooth pore model used throughout this paper. The total pore height is denoted $h$, 
    the width of the repeated unit cell is $2\Delta x=1$. The ``tooth'' heights on the top and bottom are, 
    denoted $\Delta y_t$ and $\Delta y_b$ respectively.}
  \label{pore-picture}
\end{figure}

\noi
The special class of polygons that we consider consists of channels that are periodic in the 
$x$ direction, but bounded by walls in the $y$ direction, as depicted in \fig{pore-picture}. The walls consist of straight edges, and are arranged in a saw-tooth configuration such that the top and bottom walls are ``in phase'', i.e.\ the peaks of the lower and upper walls have the same horizontal coordinate. The channel can therefore be represented as an elementary cell (EC), replicated along the $x$-axis.
We denote by $h$ the height of this cell, and set its length to $2\Delta x = 1$ \footnote{A simple
mirror symmetry operation allows the equilibrium dynamics of the billiard, defined above, to be reduced to
an even simpler fundamental domain than the EC, which is only half of the EC. However, the nonequilibrium
dynamics which will be considered later, is made of curved trajectories, whose convexity has a
precise sign, and is not preserved by the mirror symmetry. Therefore, we do not reduce further the EC.}.
The heights of the isosceles triangles comprising the ``teeth'' along the top and bottom cell walls are denoted $\Delta y_t$ and $\Delta y_b$ respectively. We also introduce the mean interior channel height $d$, defined as $d=h-(\Delta y_t+\Delta y_b)/2$, which is equal to the mean height of the pore volume accessible to particles inside the channel. For convenience, we introduce the angles
\be
\theta_i = \tan^{-1} \left( \frac{\Delta y_i}{\Delta x_i} \right) ~, \quad \mbox{for } i=b,t ~.
\ee 

In this paper we consider a range of values of $h$, $\Delta y_t$, and $\Delta y_b$, which can be classified into two groups. First, we
consider systems where $\Delta y_t=\Delta y_b$ (and consequently $\theta_t = \theta_b$). In this case, the top and bottom saw-teeth
are parallel with one another. Alternatively, we consider systems where $\Delta y_t$ and $\Delta y_b$ are unequal, and the saw-tooth
walls are not parallel. The choice of $\Delta y_t = 0, \Delta y_b\not=0$ (or vice versa) is a special case of this last group -- in
this case, the horizontal top wall induces a vertical symmetry such that the dynamics is
isomorphic to a system where both top and bottom saw-teeth have height $\Delta y_b$, but where the saw-teeth are $\Delta x$ out of
phase, as in the equilibrium mechanical pump of Ref.\cite{BRMPEJ}. We will therefore consider these walls as an extension of the
first group of parallel saw-tooth walls. 

The model described so far can be called an {\em equilibrium} model, because there is no dissipation of energy, due to external
forces or otherwise. A {\em nonequilibrium} model can be constructed introducing the influence of an external
field $\epsilon$, that accelerates the particle in the positive $x$ direction, and of a Gaussian thermostat, which balances the
effect of the field, dissipating energy in such a way that the speed of the particle becomes a constant of motion. In this case, the
particle obeys the equations of motion
\begin{equation}
\left. 
\begin{array}{rl}
  \dot{x} = p_x &\qquad \dot{p}_x = -\alpha p_x + \epsilon \quad \\
  \dot{y} = p_y &\qquad \dot{p}_y = -\alpha p_y
\end{array} 
\right\}
\quad \alpha = \epsilon p_x
\label{noneqeqmot}
\end{equation}
until it reaches the boundary $\partial {\mathcal P}$, where it undergoes a specular 
reflection, as in the equilibrium case. The effect of the field is to curve the trajectories, making
them concave in the direction of the field. In fact the solution of the equations of motion, for the 
free flight parts of the trajectory are given by: 
\bea
&&\tan{\frac{\zvt(t)}{2}} = \tan{\frac{\zvt_0}{2}} \, e^{-\epsilon t}
\label{teta}\\
&&x(t) = x_0 - \frac{1}{\epsilon} \ln {\frac{\sin \zvt(t)}{\sin \zvt_0}}\\
&&y(t) = y_0 - {\frac{\zvt(t) - \zvt_0}{\epsilon}},
\eea
which depend on the initial angle $\zvt_0$. We note that the boundary of this system is not defocussing, and the external field has a focussing effect, so that the overall dynamics should not be chaotic, although it is not obvious that this is the case for all values of $\ze$. When the external field $\epsilon$ is set to zero, one recovers the usual equilibrium equations of motion.

The data in this paper were obtained from molecular dynamics simulations of the channel transport system described above. In both
the equilibrium and nonequilibrium cases, the momentum and position of the particle are determined from analytic solutions of the
equations of motion. Reported values correspond to averages of distinct simulation runs with different initial conditions for the
particle. These initial conditions were randomly generated with a uniform spatial distribution, and a ``circle'' velocity
distribution, expressed by a probability density
\be
\frac{1}{2 \pi} \zd(\rho-1) \, d \rho \, d \zvt ~, \quad \text{with velocity } v = \rho (\cos \zvt , \sin \zvt)
\ee 
where the speed is $|v|=\rho=1$.


\section{Theory}
\label{theory}

Generally, our interest in studying ensemble properties stems from the foundations
of statistical mechanics. From a practical viewpoint, when studying a molecular system, we expect that the specific initial
microscopic conditions do not alter the thermodynamic properties of a system. That is, we expect an equivalence between the
ensemble and time averages of trajectory properties, such that if we were to wait long enough, the average of a property along
almost any trajectory would be independent of the initial condition (and equal to the average of the property on the ensemble of
initial conditions). Such a phenomenological requirement is incorporated into the theoretical structure of statistical
mechanics through the mathematical notion of ergodicity, which can be summarized as follows. Consider a particle system constituted
by $N$ classical particles, described by the equations of motion
\begin{equation}
\dot{x} 
= G (x) ~; \quad  x = ({\bf q},{\bf p})  \in \mathcal{M} \subset \zR^{6N} ,
\label{xdot}
\end{equation}
where $\mathcal{M}$ is the phase space, and the vector field $G$ contains 
the forces acting on the system, and the particles interactions. 
Denote by $S^t x$, $t \in \zR$, the solution of \q{xdot} with initial 
condition $x$. The macroscopic quantity associated with an \emph{observable}, 
i.e.\ with a function of phase $\zF : \zW \rightarrow \zR$, is defined by:
\begin{equation}
\bar{\zF}(x) = \lim_{T\rightarrow\infty} \frac{1}{T} \int_0^T \zF(S^t x) d t  ~.
\label{Taverage}
\end{equation} 
in which the time average represents the fact that macroscopic observations 
occur on time-scales which are very long compared to the microscopic time-scales,
so that the measurement amounts to a long time average. In general, however, 
computing that limit is not a trivial task at all. The problem is commonly solved 
by invoking the EH, which states that 
\begin{equation}
\bar{\zF}(x) = \frac{1}{\mu (\Omega)} \int_\zW \zF(y)  \, d \mu (y) 
= \langle \zF \rangle_{\mu}                         
\label{EH}
\end{equation}
for a suitable measure $\mu$ (the {\em physical measure}), and for $\mu$-almost all 
$x \in \Omega$. In principle, ergodic theory should identify the cases verifying 
the EH, and the physical measures $\mu$, but in practice, this is too hard, if not
impossible, to do. Nevertheless, there is 
now a vast literature on the validity of the EH, which can be understood
in different ways, but which finds in Khinchin's arguments on the properties of sum
variables, the most convincing explanation \cite{Kinchin}. In practice, the time average
of the functions of physical interest in systems of many particles is reached before a 
trajectory has explored the whole phase space (which would take too long), because
such functions are almost constant, and equal to the ensemble average. Therefore, the
finer details of the microscopic dynamics are not particularly important, except for the
requirement of some degree of ``randomness'' (in order to introduce a decay in correlations
between particles in the system). However, it is not clear what properties should be
imposed on the dynamics in order to obtain sufficient randomness. Consequently, in a system devoid
of dynamical chaos, such as that under investigation in this paper,
it is of interest to investigate the behaviour of both the
individual and ensemble properties of particles (and their trajectories). 

From a thermodynamic viewpoint, diffusion is the transport process generated by gradients in the
chemical potentials of the different chemical species in multicomponent thermodynamic systems,
or of tagged particles moving in a host environment made of (mechanically) identical
particles (self diffusion) \cite{DGM,CC70}. More commonly (and often more conveniently), this process is described
in terms of Fickian diffusion, relating the mass flux to gradients in local density. Through the
fluctuation-dissipation relation (which rests on the EH), the same diffusion properties are also related to the relaxation
of local mass-gradient fluctuations at equilibrium, and therefore to the manner in which the
positions of particles evolve during equilibrium dynamics, although the response to an external
action, and the spontaneous equilibrium fluctuations are rather different phenomena 
\footnote{For instance, the spontaneous fluctuations around equilibrium states do not dissipate any
  energy. Indeed, they do not change the state of the system, and continue forever. Differently, the
  response to an external action dissipates part of the energy received, and may modify the state of
  the system, or maintain a nonequilibrium steady state.}. 
Fick's first law for diffusion is expressed by
\cite{DGM}:
\be
{\bf J}({\bf x}) = -D \nabla n ({\bf x})
\ee
where ${\bf J}$ is the mass flow, $D$ is the (Fickian) self diffusion coefficient, $n$ is the number density, and
{\bf x} is the position in space. This law, which can be justified in kinetic theory \cite{CC70}, provides the 
phenomenological basis for the mathematics of diffusion in molecular systems, leading to the second-order PDE 
\begin{equation}
\label{fick}
\dd{n}{t} = D \dd{^2 n}{x^2}.
\end{equation}
known as Fick's second law, where $t$ is the time variable. The well-known Gaussian evolution
\begin{equation}
\label{gaussian}
n(x,t) = (4\pi Dt)^{-1/2} e^{-x^2/4Dt}
\end{equation}
results from an initial delta-function distribution, and the linearity of \q{fick} ensures that the diffusion of 
a system of molecules can be considered as the evolution of a superposition of Gaussians. In particular, we recover 
from \q{gaussian} the linear growth in the mean-square displacement for macroscopic diffusion processes 
\begin{equation}
\label{msd}
\ave{x^2(t)} = \int_{-\infty}^{\infty} x^2 n(x,t) \d{x}  = 2Dt.
\end{equation}
We note that, if $n(x,t)$ is not a slowly-varying function, higher-order corrections may be introduced. 
The next approximation has the form 
\begin{equation}
\label{fick2}
\dd{n}{t} = D \dd{^2 n}{x^2} + B \dd{^4 n}{x^4},
\end{equation}
where $B$ is called the \emph{super Burnett} coefficient. In this case, the diffusion coefficient
can still be defined as in \q{msd} --- furthermore, the super Burnett coefficient can be determined
via the relation
\begin{equation}
\label{superBurnett}
\ave{x^4(t)} - 3\ave{x^2(t)}^2 = \int_{-\infty}^{\infty} x^4 n(x,t) \d{x} - 3 \left[ \int_{-\infty}^{\infty} x^2 n(x,t) \d{x} \right]^2 = 24Bt.
\end{equation}
As such, the super Burnett coefficient can be seen as a measure of the degree to which transport is
diffusive, in the Fickian sense. We note, however, that it has become customary to call {\em
diffusive} any phenomenon displaying a linear relation between the mean square displacement and the
time, like \q{msd}, even if there are no multicomponents, or any kind of particle systems at
hand. In this paper, we do the same. However, we note the key role played by the assumption of a
phenomenological law, such as Fick's, in the preceding argument. In general, for a differing
phenomenology, one cannot expect the resulting transport processes to remain diffusive in nature. In
the absence of intermolecular interactions (or indeed, of other molecules), there is no
phenomenological basis for expecting the mass flux to depend on density gradients, and it is
therefore of interest to examine the resulting transport. 

In characterising the transport law, we consider the behaviour of the displacement as a function of the
time $t$, denoted $s_x(t)$, at equilibrium. In general, we expect an (asymptotic) relation of the form
\begin{equation}
\label{defn:gamma}
\ave{s_x^2(t)} \sim A t^\gamma
\end{equation}
where the coefficient $A$ represents a mobility, and the exponent $\gamma$ indicates the
corresponding transport law. The symbol $\langle \cdot \rangle$ indicates an ensemble average, which
in equilibrium systems is universally derived from the equal {\em a priori} probability assumption,
or the EH. In the following we adopt the same assumption, when considering
equilibrium systems, although it is not obvious that one should necessarily do so. Similarly, we
adopt the language of thermodynamics, and define the following transport properties:

\vskip 5pt \noi
{\bf Definition 3. }{\it Assume that $\lim_{t \to \infty} \ave{s_x^2(t)} /
t^\zg = A$, for some  $A \in(0, \infty)$, then
\begin{itemize}
\item[{\bf i.}] {\it
If  the exponent $\zg$ equals 1, the transport is called {\em diffusive};}
\item[{\bf ii.}] {\it If $\gamma>1$, the transport is called \emph{super-diffusive} and, in particular, 
it is called  \emph{ballistic} if $\gamma=2$ (the mean square displacement is proportional to time);
\item[{\bf iii.}] The transport is called \emph{sub-diffusive} if $\gamma<1$.}
\end{itemize}
}
\noi
Away from equilibrium, it is less straightforward to distinguish between the various transport laws,
as the thermostat imposes an upper limit of linear growth for $s_x$, even for super-diffusive
transport processes. However, for the diffusive process, we expect the diffusion coefficient,
estimated from nonequilibrium transport for a given external field of strength $\epsilon$, to
converge to a well-defined value in the zero-field limit. To make this explicit, we define the
quantity
\begin{equation}
\label{defn:D-eps}
D(\epsilon) = \frac{kT\ave{v_x}}{m\epsilon},
\end{equation}
for a system of particles of mass $m$ at temperature $T$ (with Boltzmann's constant $k$), and mean streaming velocity of $\ave{v_x}$, as the \emph{finite-field estimate}, at field $\epsilon$, of the (nonequilibrium) diffusion coefficient. The existence of a linear regime guarantees the convergence of $D(\epsilon)$, in the zero-field limit, to a diffusion coefficient $D$ --- that is, 
\begin{equation}
\label{defn:D-noneq}
D = \lim_{\epsilon \rightarrow 0} D(\epsilon) = \frac{kT}{m} \lim_{\epsilon \rightarrow 0} \frac{\ave{v_x}}{\epsilon}
\end{equation}
Thus, for a \emph{diffusive} process, we expect to observe a linear response to the external field. Furthermore, in a thermodynamic system, we expect this diffusion coefficient to be equal to that describing equilibrium diffusion. For a super-diffusive process, however, we do not expect a linear response, but rather that $D(\epsilon)\rightarrow\infty$ in the zero-field limit. Due to the thermostat, the maximum mean value of $v_x$ is equal to the initial speed of the particle $v$. For initial velocity set to unity, $D(\epsilon)$ has upper bound of $1/2\epsilon$ (in reduced units) for any $\epsilon$, so that $D(\epsilon)$ can only diverge in the zero-field limit.


\section{Results}
\label{results}
In this section we outline the results we have obtained from simulations of the transport of molecules in the saw-tooth channel system described in section \ref{system}. We examine the equilibrium transport properties in section \ref{equilibrium}, and the nonequilibrium transport properties in section \ref{nonequilibrium}. In both cases, we will be interested in the collective, ensemble behaviour of particles in the system, and how this ensemble behaviour relates to the behaviour of individual trajectories.

\subsection{Equilibrium}
\label{equilibrium}

The ergodic properties of our systems are not obvious, therefore, there seems to be no immediate 
choice for a probability distribution in phase space, to be used for the ensemble averages.
Nevertheless, the uniform probability distribution (Lebesgue or Liouville measure, defined above)
is invariant, and one could think that it is appropriate for transport in a membrane which receives
particles from a reservoir, inside which the dynamics is chaotic. 
Therefore, the ensemble averages in this section are all 
computed assigning equal weight to all regions of phase space.

\subsubsection{Parallel walls, collective behaviour}

\begin{figure}
  \centering
  \epsfig{file=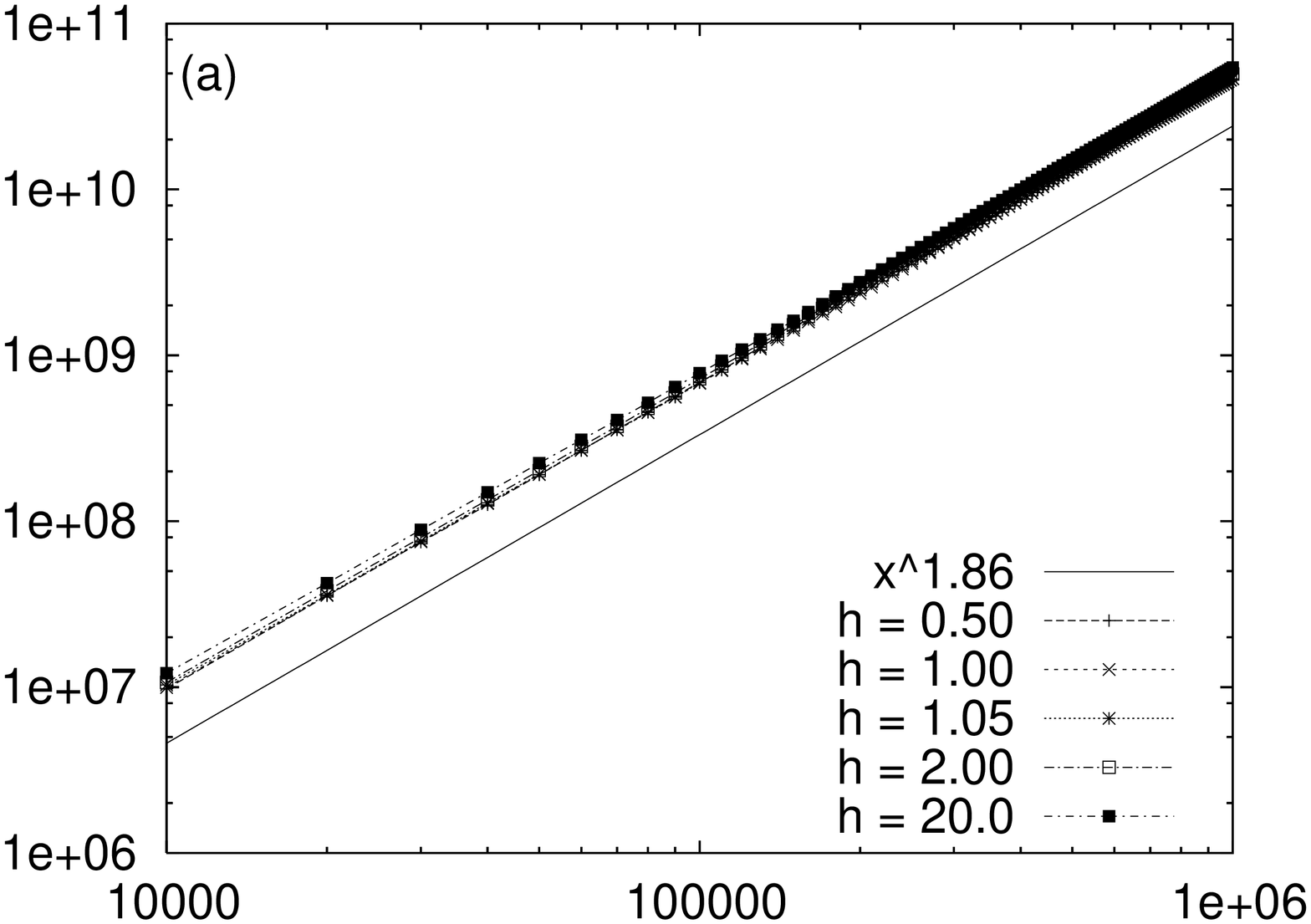 , angle=-90 , width=7cm }
  \epsfig{file=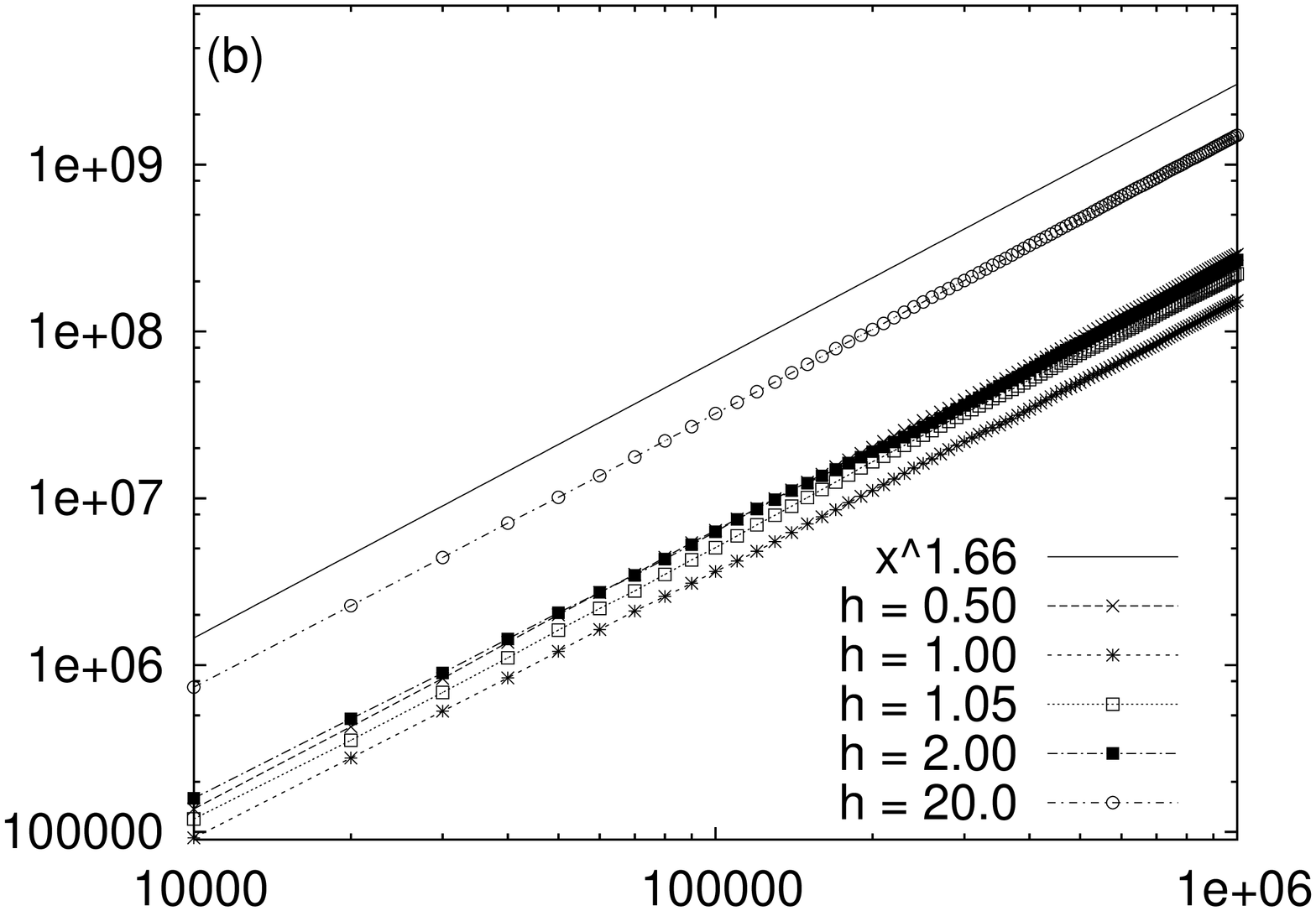 , angle=-90 , width=7cm }
  \epsfig{file=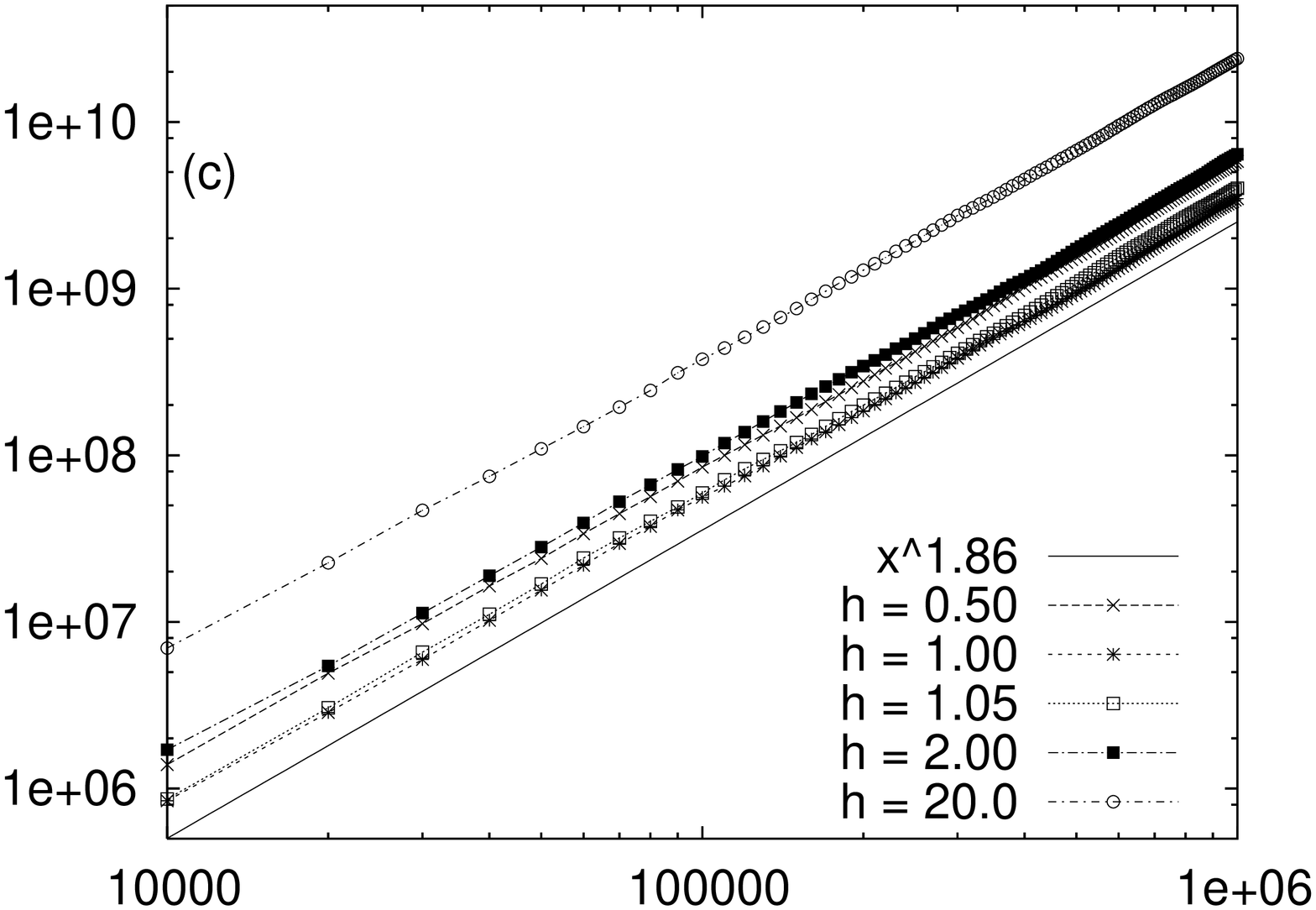 , angle=-90 , width=7cm }
  \epsfig{file=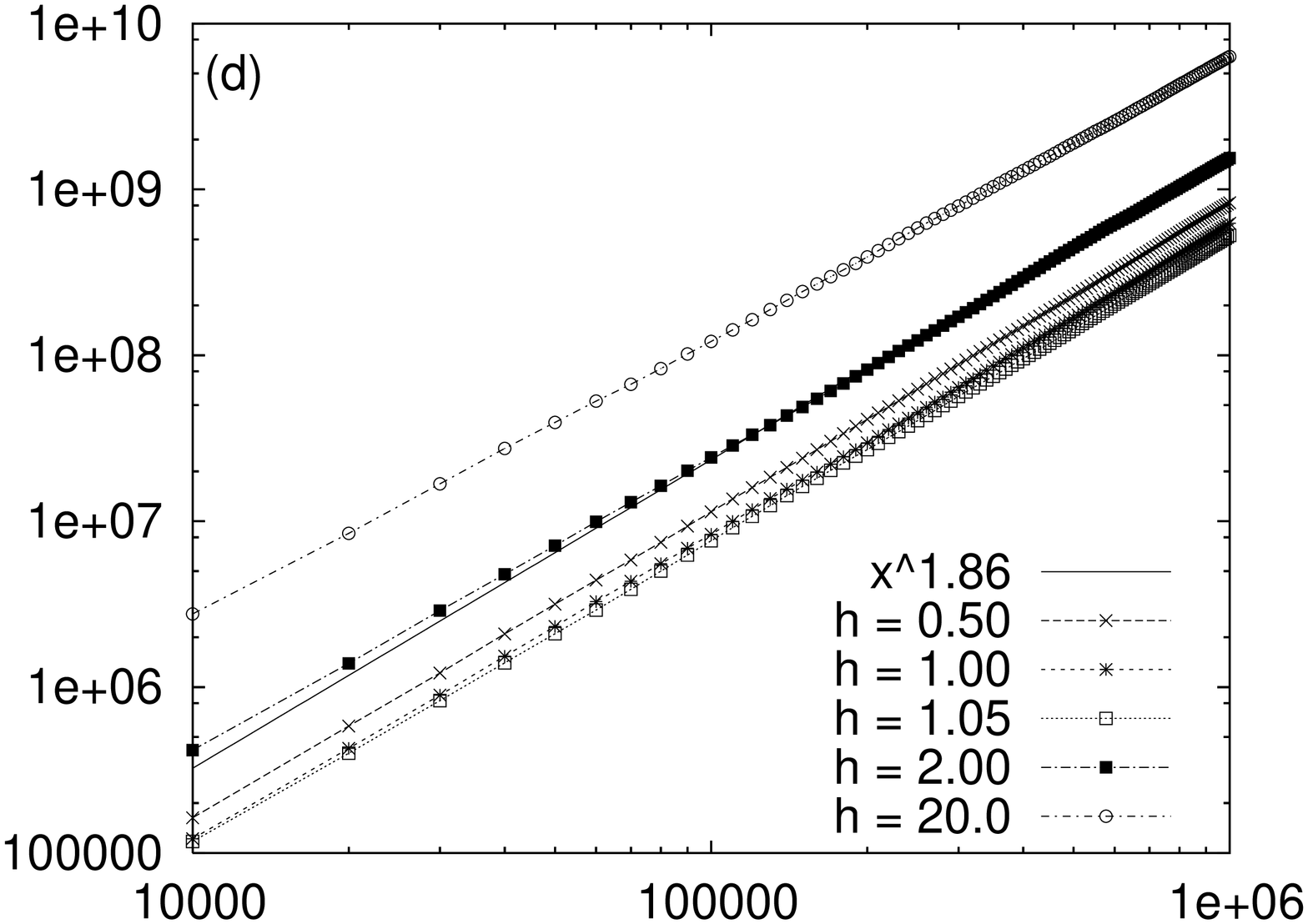 , angle=-90 , width=7cm }
  \caption{Evolution of mean-squared displacement for parallel saw-tooth systems, for 
  (a) $\Delta y/\Delta x=0.25$, (b) 1, (c) 2, and (d) 3. The data are obtained from 1000-5000
  initial conditions, and the average number of collisions is of the same order as the total time.}
  \label{par-msd}
\end{figure}

In \fig{par-msd} we depict the mean-square displacement, as a function of time, for a
series of parallel saw-tooth systems ($\Delta y_t = \Delta y_b = \Delta y$), and we
examine systems where the ratio $\Delta y/\Delta x$ varies from 0.25 to 3, so that the
angle $\theta=\theta_t=\theta_b$ the saw-tooth makes with the horizontal varies from about
$0.08 \pi$ radian ($\approx14^\circ$) to about $0.4\pi$ radians ($\approx72^\circ$). Each graph
shows results for a single value of $\Delta y/\Delta x$, for various pore heights $h$. For
each choice of $\Delta y$ (recalling that $\Delta x = 0.5$), we examined pores with
heights $h = 1.5\Delta y, 2\Delta y, 2.05\Delta y, 3\Delta y$ and $21\Delta y$. The
corresponding interior pore heights are $d=0.5\Delta y, \Delta y, 1.05\Delta y, 2\Delta y$
and $20\Delta y$. We note that $h=2\Delta y$ corresponds to the critical pore width, above
which the billiard horizon is infinite (i.e.\ there is no upper bound to the length of
possible molecule trajectory segments without boundary collisions). 
We considered sets of initial
conditions ranging from 1000 to 5000 particles.
Not surprisingly, given the clearly non-diffusive nature of the transport observed in
\fig{par-msd}, the corresponding super Burnett coefficients do not appear to have a well-defined value, but
diverge over time.

The same data have also been generated for a series of systems with one flat wall and one saw-tooth
wall, which always have an infinite horizon. For these systems, we set $\Delta y_t = 0$, and
considered the same ratios $\Delta y_b/\Delta x$ as were examined in the parallel saw-tooth systems
with pore heights $h = 1.05 \Delta y_b, 2.5 \Delta y_b$ and $20.5 \Delta y_b$ (which have corresponding
interior heights $d = 0.55 \Delta y, 2 \Delta y$ and $20 \Delta y$). Sets of initial conditions
ranged from 1000 to 5000 particles.  There appeared to be a longer initial transient period for
the systems with one flat wall, but otherwise the results were qualitatively the same as for the
parallel wall, despite the infinite horizon.

\tabl{par-flat} shows the values of the exponents obtained from fitting the data in \fig{par-msd}
and those for one flat wall to \q{defn:gamma}, for both the parallel saw-tooth systems and the systems with
one flat wall and one parallel wall. In all cases, we observe that the transport is significantly
super-diffusive, but \emph{not} ballistic. For the $\Delta y/\Delta x=1$ system, which is a
(parallel) rational polygonal billiard, we observe an exponent $\gamma\approx1.65$. All other
choices of $\Delta y/\Delta x$ represent (parallel) irrational polygonal billiards. For $\Delta
y/\Delta x=1/4$ (where $\theta<\pi/4$), we observe an exponent of $\gamma\approx1.85$ in all
systems. For $\Delta y/\Delta x=2$ and $\Delta y/\Delta x=3$ (where $\theta>\pi/4$), we observe a
similar exponent in the systems with pore height less than or in the vicinity of $2 \Delta y$,
and a \emph{reduction} in the transport exponent as the pore height increases.
The same value $\gamma \approx 1.85$ has been found
by M. Falcioni and A.Vulpiani in the equilibrium version of the periodic Ehrenfest gas of
Ref.\cite{jsp099_0857}, \cite{FVpriv}.

\begin{table}
  \centering
  \begin{tabular}{c|c|c|c|c|c|c|c|c|}
    $\Delta y/\Delta x$ & \multicolumn{5}{c|}{saw-tooth systems} &\multicolumn{3}{c|}{saw-tooth/flat systems} \\
                        & $0.5 \Delta y$  & $1.0 \Delta y$  & $1.05 \Delta y$ & $2.0 \Delta y$  & $20 \Delta y$   & $0.55 \Delta y_t$ & $2.0\Delta y_t$ & $20 \Delta y_t$ \\
\hline
            0.25        &       1.85      &       1.83      &       1.82      &       1.85      &       1.85      &       1.88        &      1.83       &       1.85       \\
            1           &       1.66      &       1.64      &       1.62      &       1.67      &       1.68      &       1.65        &      1.65       &       1.65       \\
            2           &       1.83      &       1.85      &       1.82      &       1.80      &       1.79      &       1.83        &      1.82       &       1.75       \\
            3           &       1.86      &       1.87      &       1.84      &       1.80      &       1.70      &       1.82        &      1.76       &       1.70       \\
  \end{tabular}
  \caption{Equilibrium transport exponents: For saw-tooth boundary base triangles with height-width ratio $\Delta y/\Delta x$. For each (mean) pore height tested, the observed exponent out to $10^6$ time units (of the order of $10^6$--$10^7$ collisions) is given. The number of initial conditions used to compute averages ranges from 1000 to 5000, and the errors are estimated by $\pm 0.03$ in all cases.}
  \label{par-flat}
\end{table}

\begin{figure}
  \centering
  \epsfig{file=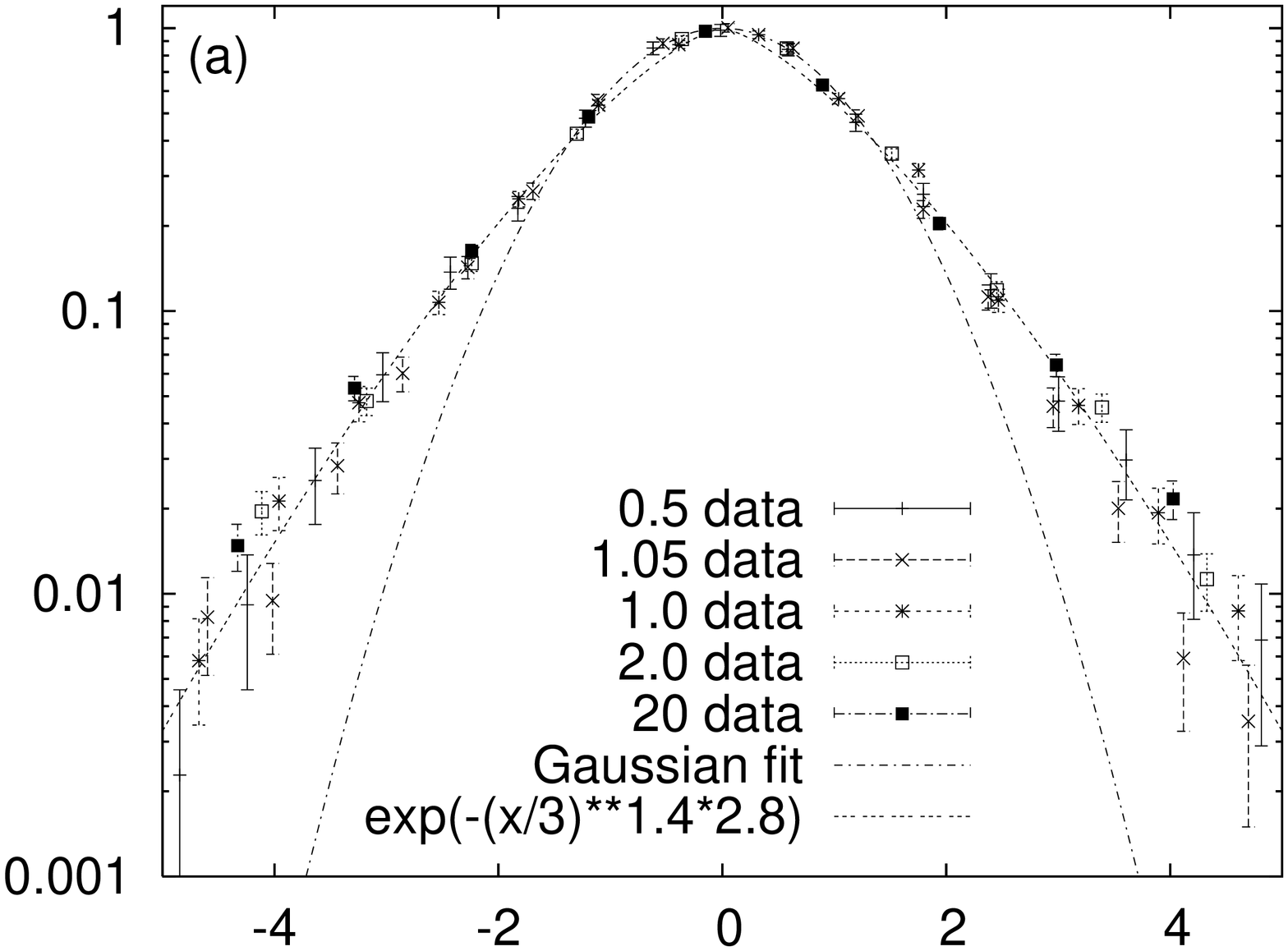 , angle=-90 , width=8cm }
  \epsfig{file=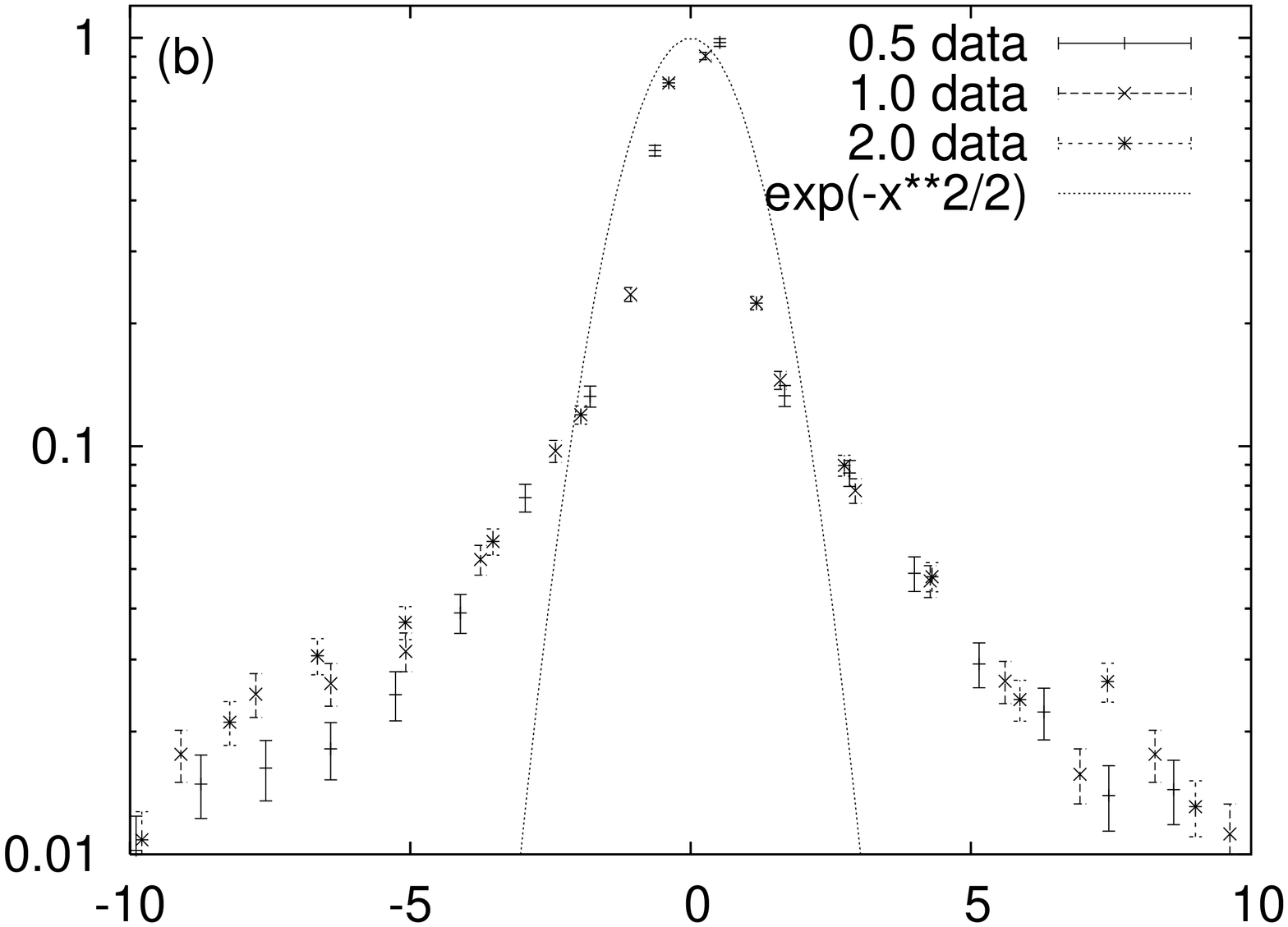 , angle=-90 , width=8cm }
  \caption{Distribution of displacements after $10^6$ time units for parallel saw-tooth systems, where (a) $\Delta y/\Delta x=2$, from 2000 initial conditions, and (b) $\Delta y/\Delta x=1$, from 5000 initial conditions. }
  \label{par-dists}
\end{figure}

We have also examined the distribution of the total $x$ displacements $s_x(t)$, as a function of
time. Distributions for the $\Delta y/\Delta x=1$ system (obtained from 2000 initial conditions) and
the $\Delta y/\Delta x=2$ system (obtained from 5000 initial conditions), obtained at the
end of the simulations, are shown in \fig{par-dists}. The results for the $\Delta y/\Delta x=2$
system are typical of the results observed for the other parallel irrational polygonal billiards. Errors are
estimated from the frequency counts used to generate the histograms. 

From an initial distribution that is effectively a delta function on the scale of the motion, the
transport process produces symmetric distributions, distinct from a Gaussian distribution. For the
parallel irrational polygonal billiards, the distribution appears Gaussian out to one standard
deviation --- however, the distribution at larger displacements is clearly underestimated by the
Gaussian. Attempted fits using functions of the form $\exp\{-x^{\alpha}\}$ fail to capture the shape
both at the centre and in the tails, although the tails appear to be well modelled by a distribution
of the form $\exp\{-x^{1.4}\}$ (\fig{par-dists}a). When $\theta = \pi/4$, however, the distribution 
bears little resemblance to a Gaussian (\fig{par-dists}b). Analogous results in each case were 
obtained from the systems with one flat wall and one saw-tooth wall.

\begin{figure}
  \centering
  \epsfig{file=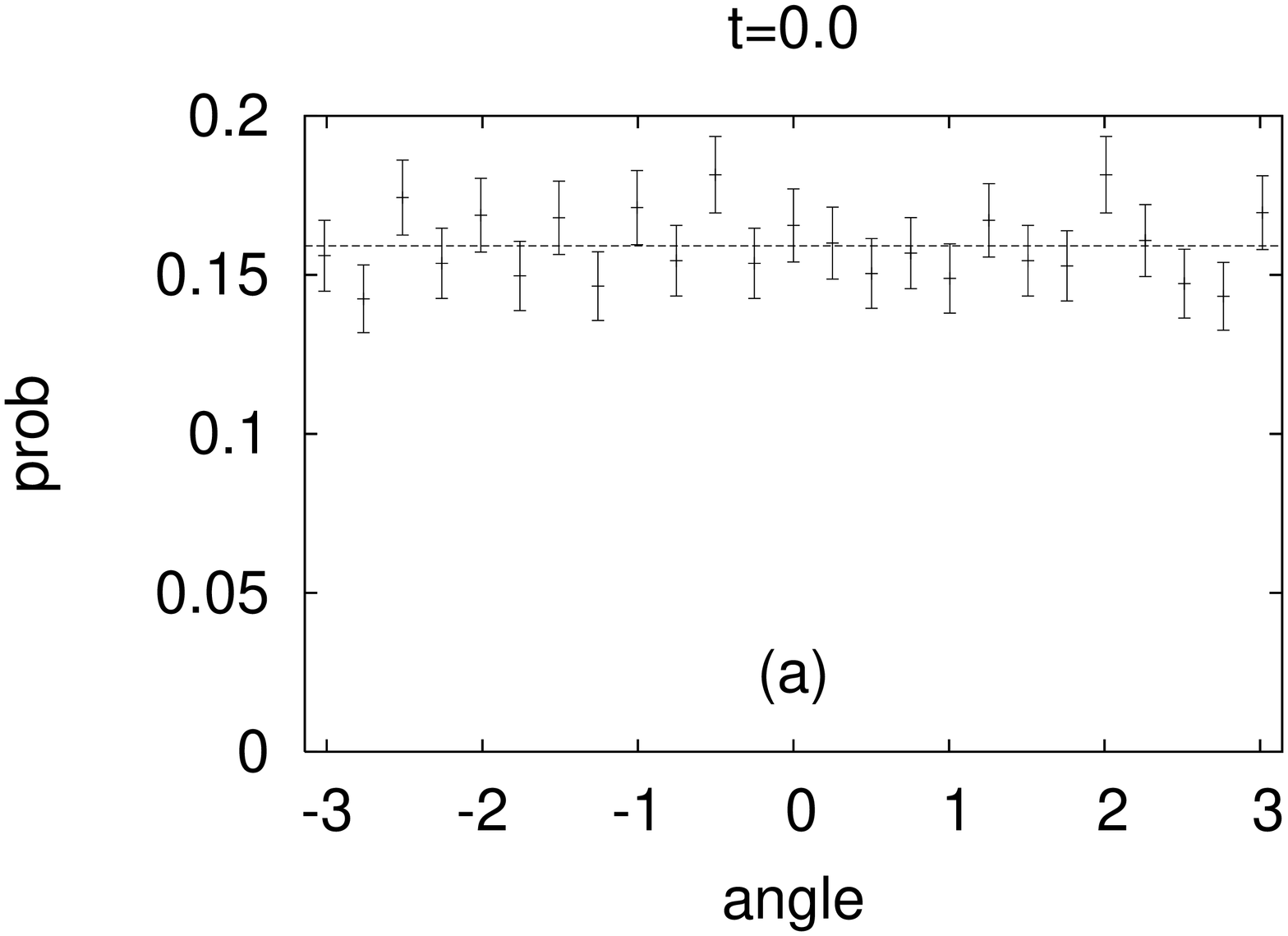 , angle=-90 , width=7cm }
  \epsfig{file=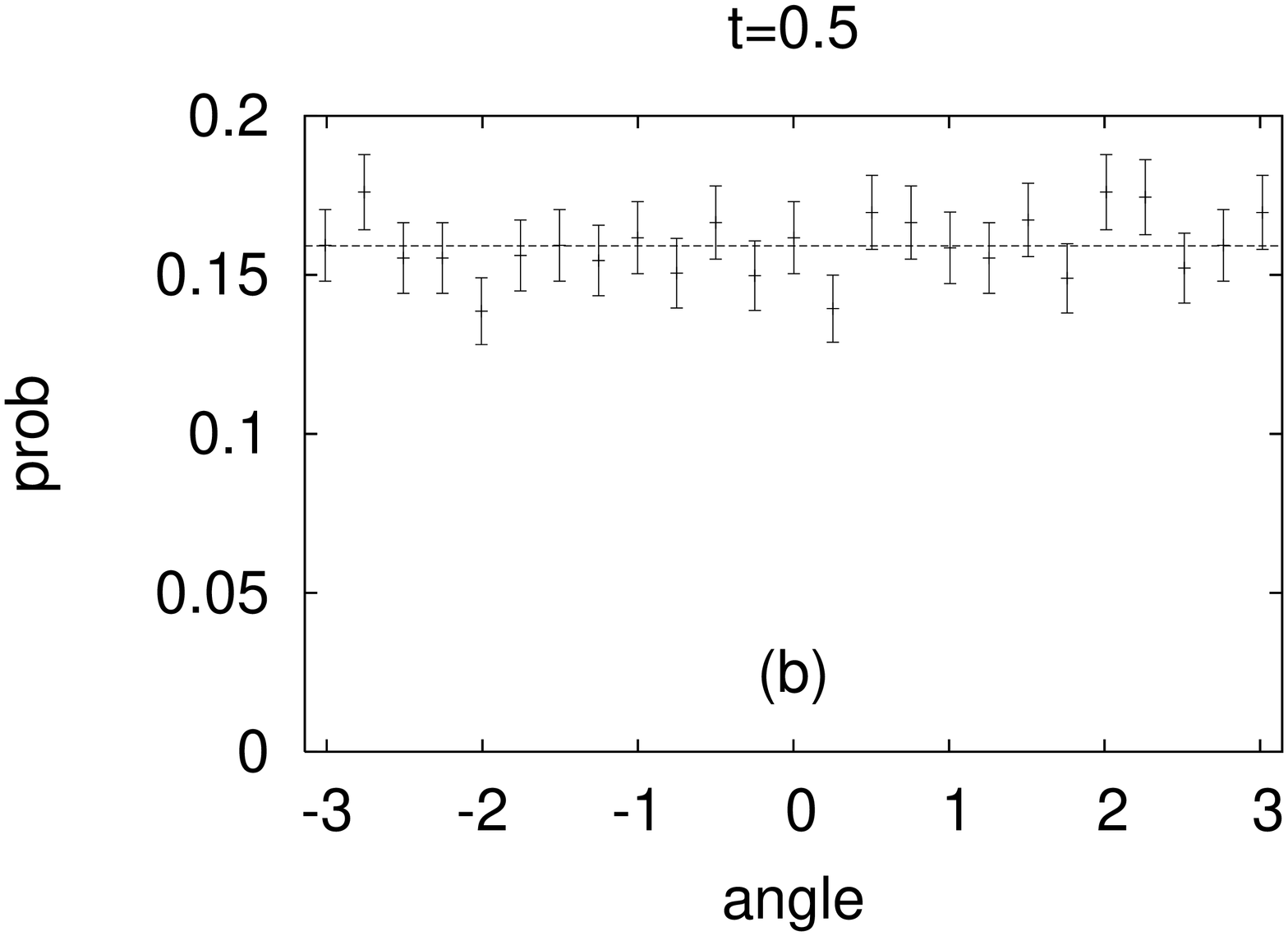 , angle=-90 , width=7cm }
  \epsfig{file=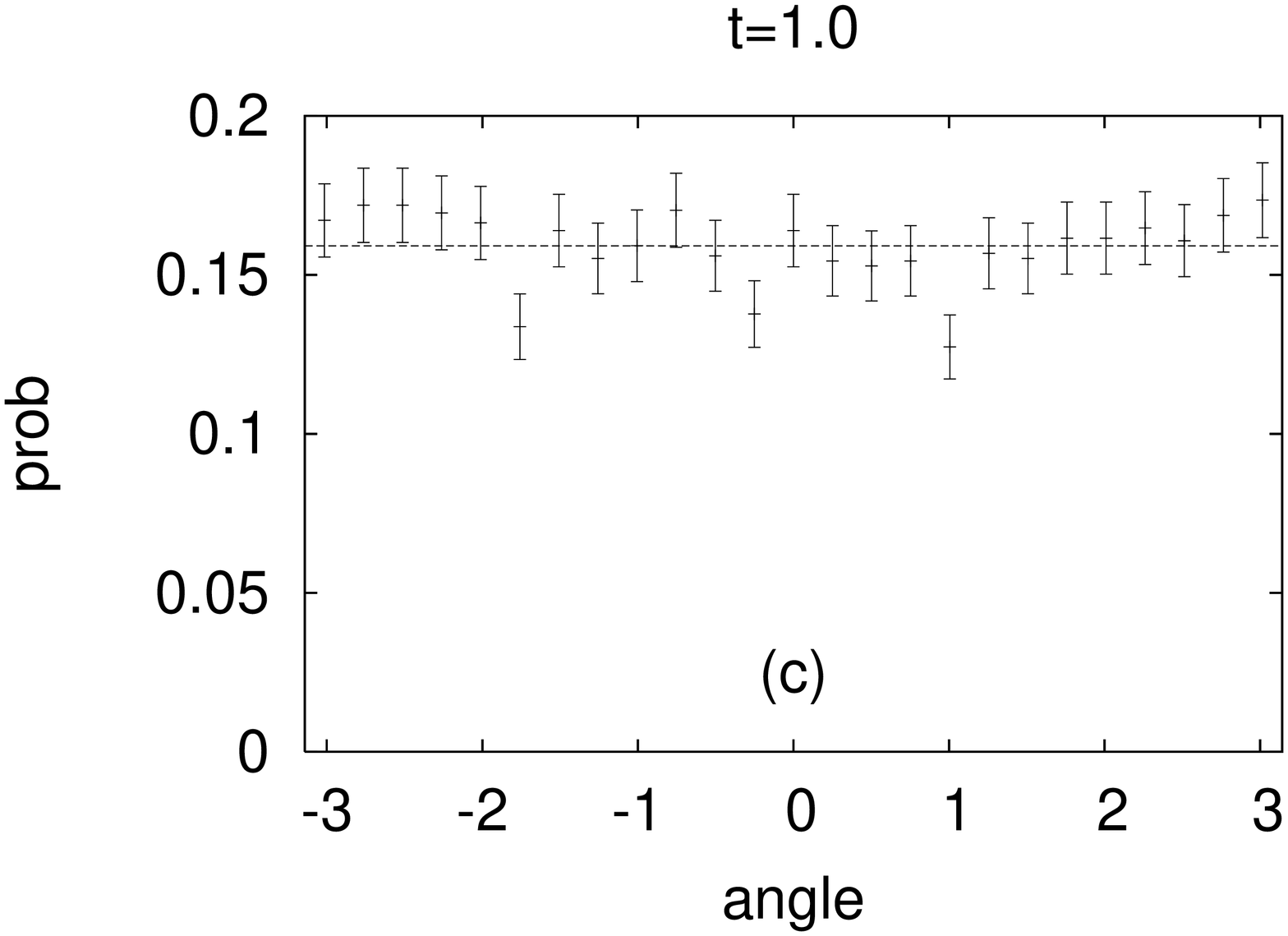 , angle=-90 , width=7cm }
  \epsfig{file=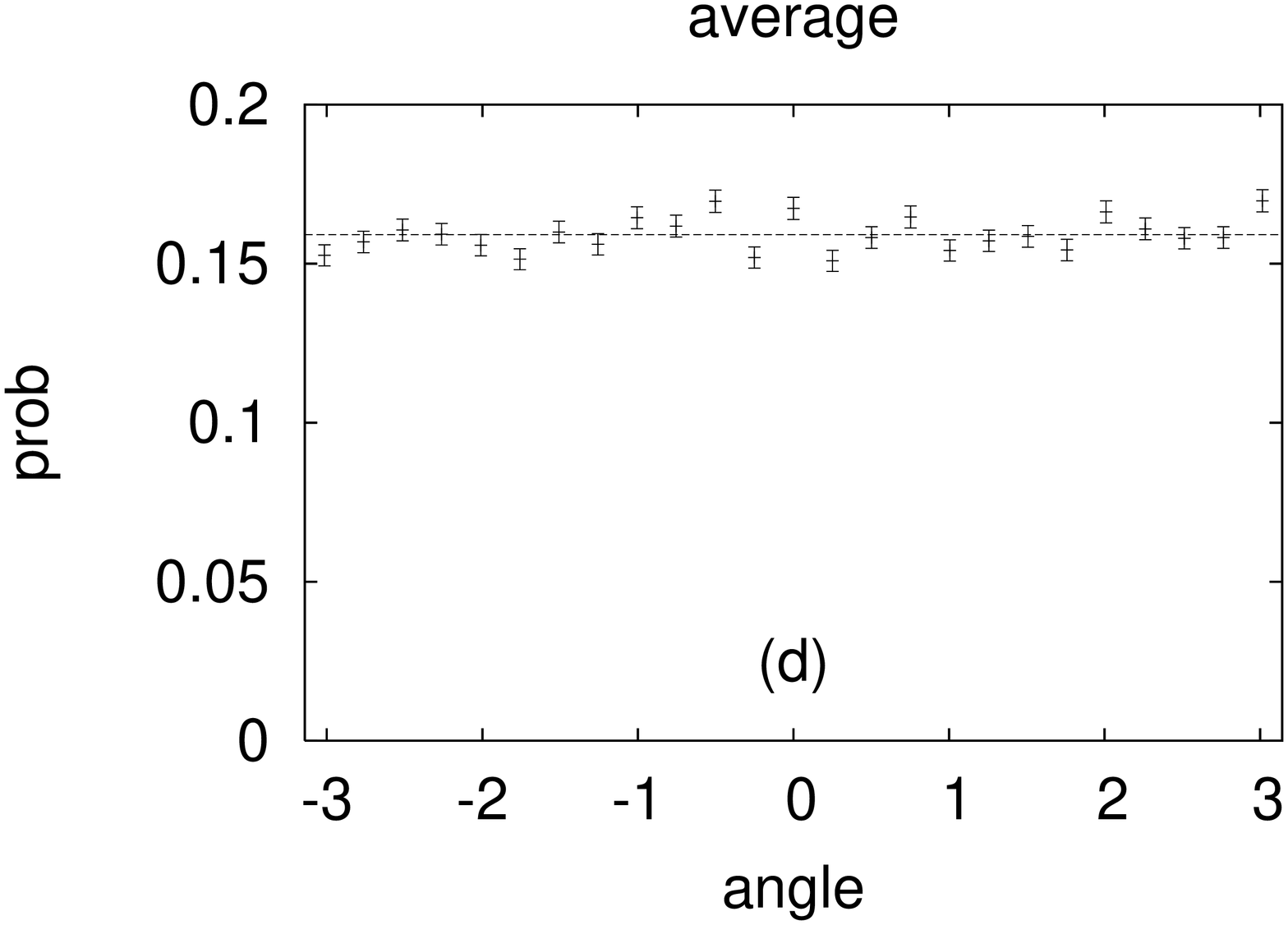 , angle=-90 , width=7cm }
  \caption{Distribution of momenta orientations (a) at $t=0$, (b) at $t=5\times10^5$, (c) at $t=10^6$, and (d) averaged over all times, for the $\Delta y/\Delta x=1$ system with $d=0.5\Delta y$. The dotted line indicates the uniform distribution. Errors are estimated from the frequency counts used to generate the histograms (and are hence smaller in (d), where the data is comprised of more samples).}
  \label{par-distp1}
\end{figure}

We have also examined the behaviour of the momenta of particles in our systems. Given that
the speed is preserved by the dynamics, the momenta can vary only in orientation, and we
therefore examine the effect of the dynamics on the distribution of these orientations. 
\fig{par-distp1}
shows the typical behaviour of the distribution of momenta orientations at the
beginning, at the midpoint, and at the end of a typical simulation of 5000
particles. We also show a mean distribution obtained from averaging the momentum data over
all sampled times, as well as over all trajectories. We find that there are no significant
correlations in the distributions of the momenta over the course of the simulation ---
while the distribution of orientations, as a function of time, does not converge to the
uniform distribution over the time-length we have considered (as we might expect it to for
a large system of interacting particles), any memory effects do not appear to have a
significant influence on the overall distribution at any instant, which remains close to
uniform. Again, the same conclusion can be drawn from similar examination of the systems
with one flat wall and one saw-tooth wall.

\vskip 5pt \noi
{\bf Remark 1. }{\it 
That the angle of the walls with the horizontal plays a role in the determination of the 
transport law, while the width of the pores does not appear to do so, is reminiscent of
the case of dispersing billiards. There, the infinite horizon adds a logarithmic correction
to the dependence of the mean square displacement on time, which is rather hard 
to detect numerically. Despite the lack of chaos, our non-dispersing billiards could enjoy
similar logarithmic corrections when the horizon is infinite. More surprising is the fact
that in some cases the transport exponent (not just the matter transport) \emph{decreases}
as the pore width increases while the horizon is
finite, and only increases again once the horizon is infinite.}

\subsubsection{Parallel walls, individual behaviours}

From the results we have obtained, there appear to be well-defined collective behaviours that are attributable to the systems we have studied --- that is to say, the mean values of the properties obtained from simulation appear to converge, in the limit of a large number of independent trajectories (or 
particles), to well-defined values. In analogy to what we have presented above, we examine the evolution of 
the individual particle momenta, and the $x$ displacements.

\begin{figure}
  \centering
  \epsfig{file=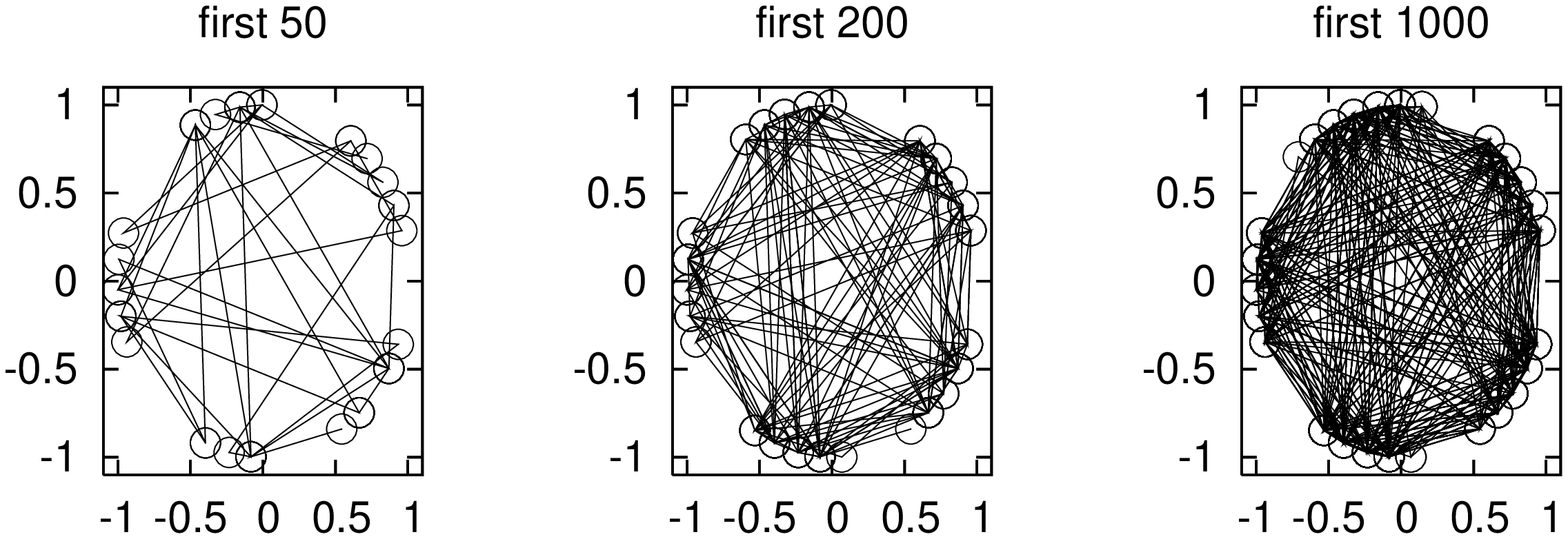 , angle=-90 , bb = 150 50 454 770,  width=12cm }
  \caption{Momentum progression for a single trajectory up to (a) $t=5\times10^5$ (50 samples), (b) $t=2\times10^6$ (200 samples), and (c) $t=10^7$ (1000 samples), for the $\Delta y/\Delta x=3$ system with $d=2\Delta y$.}
  \label{par-1p}
\end{figure}

The picture of the momenta is trivial in the $\Delta y/\Delta x=1$ systems, because for $\theta = \pi/4$ 
only four orientations per trajectory at most are possible, as determined by the initial
condition. The picture of the momenta for irrational systems is less predictable. Over the course of the
simulations, sequences of momenta, {\em sampled at intervals of $10^4$ time units}, were collected. 
In \fig{par-1p} we show such a sequence of momenta, sampled from a system where $\Delta y/\Delta x=3$, accumulated up to the $5\times10^5$ time units (\fig{par-1p}a), $2\times10^6$ time units (\fig{par-1p}b),  and $10^7$ time units (\fig{par-1p}c). The momenta are represented by symbols (circles) on the unit circle $\zT^1$, while the lines indicate the sequence of the sampled momenta. It is clear from \fig{par-1p} that, despite consisting of up to 1000 different sampled momenta, the number of distinct momenta visited by the particle is relatively small (of the order of 10-20). Furthermore, the growth of this set is gradual, and strongly correlated to the set of momenta that precede it in the sequence of momenta visited by the particle. The choice of $\theta$ irrationally related to $\pi$ permits, in principle, the exploration of the whole unit circle of orientations. However, it is clear that the nature of the sequence of wall collisions limits the rate at which such an exploration of the unit circle can be achieved. This slow growth was observed for simulation times up to $10^9$ time units (not shown here).

\begin{figure}
  \centering
  \epsfig{file=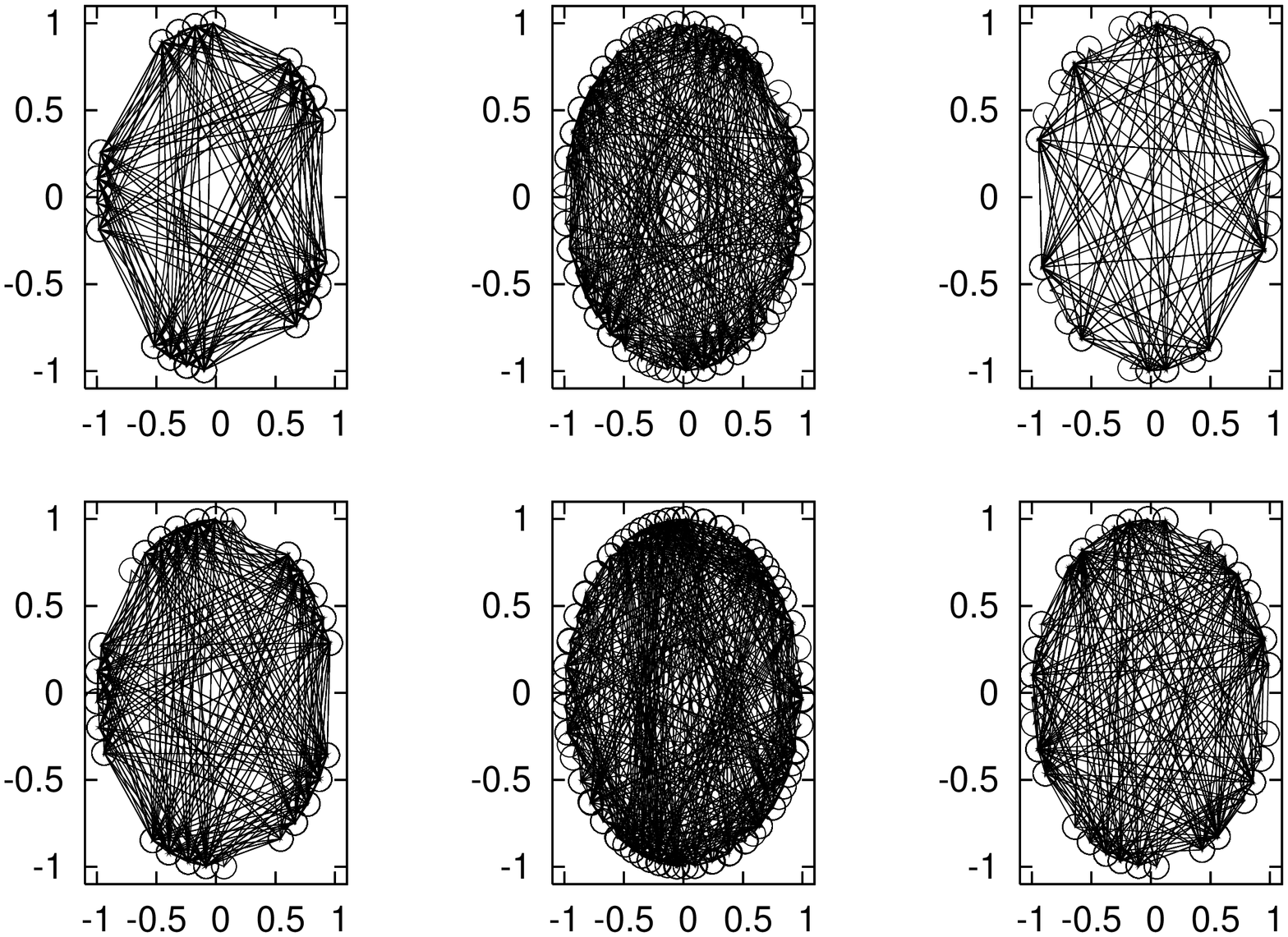 , angle=-90 , width=12cm }
  \caption{Sequences of sampled momenta for six different initial conditions, up to $10^7$ time units (1000 samples), for the $\Delta y/\Delta x=3$ system with $d=2\Delta y$.}
  \label{par-6p}
\end{figure}

In \fig{par-6p} we show the sequence of the first 1000 momenta, sampled every $10^4$ time units, for six distinct
initial conditions in a $\Delta y/\Delta x=3$ system. In each case, the available velocity phase space is gradually
explored by the particle. We note that the rate and manner in which this exploration takes place (as
indicated by the lines joining consecutive sample momenta) depends significantly, and {\em unpredictably},
on the initial condition, as is demonstrated by the visibly different structures generated by
each. This feature is common to all parallel irrational polygonal billiards examined --- for systems
with parallel saw-tooth walls and systems with one flat wall and one saw-tooth wall.
The flat wall in these latter systems induces a vertical symmetry that is absent in the parallel
systems, and appears to increase the range of momenta visited by particles --- however, a similar
degree of connectivity is observed in both cases.

\begin{figure}
  \centering
  \epsfig{file=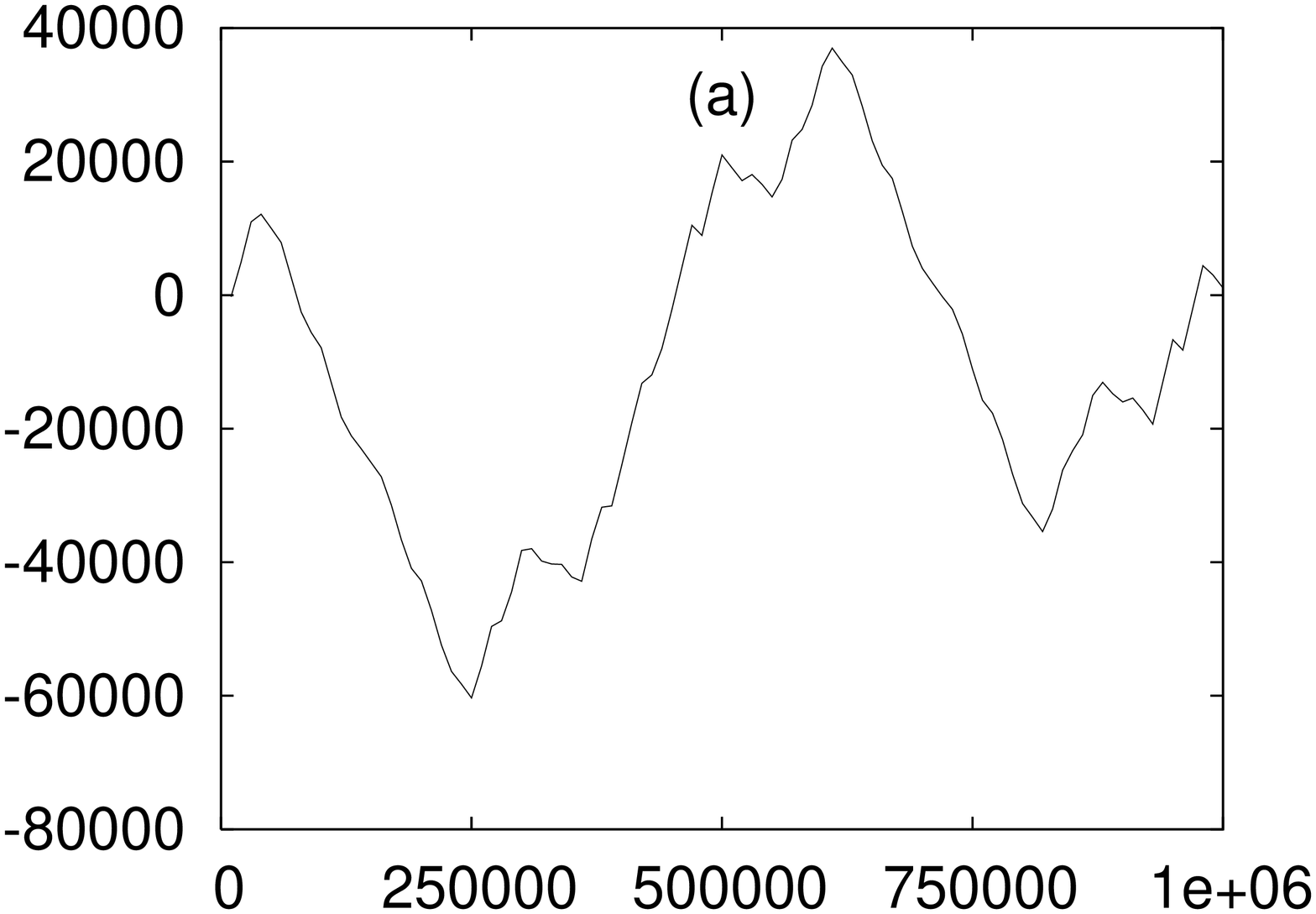 , angle=-90 , width=7cm }
  \epsfig{file=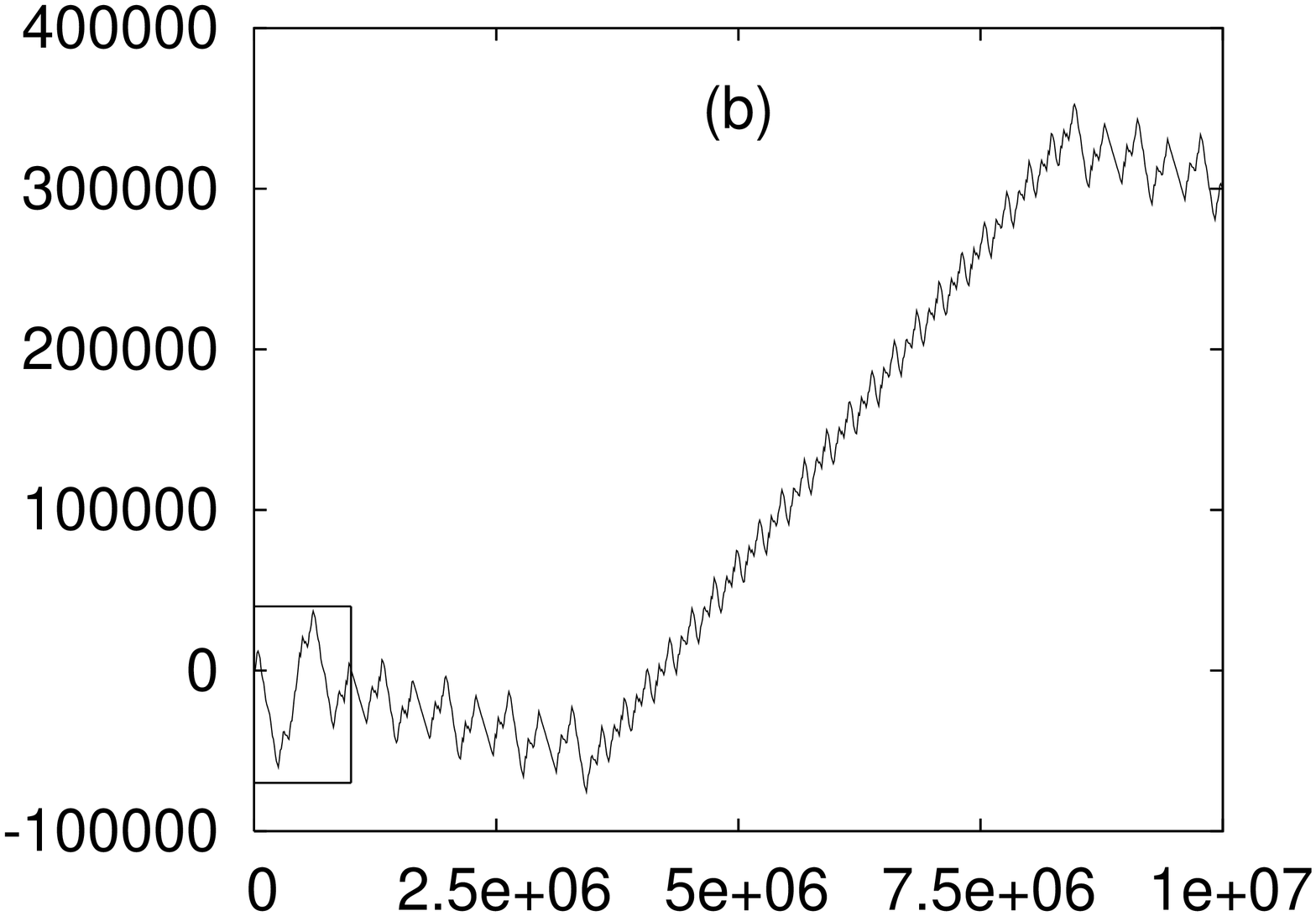 , angle=-90 , width=7cm }
  \epsfig{file=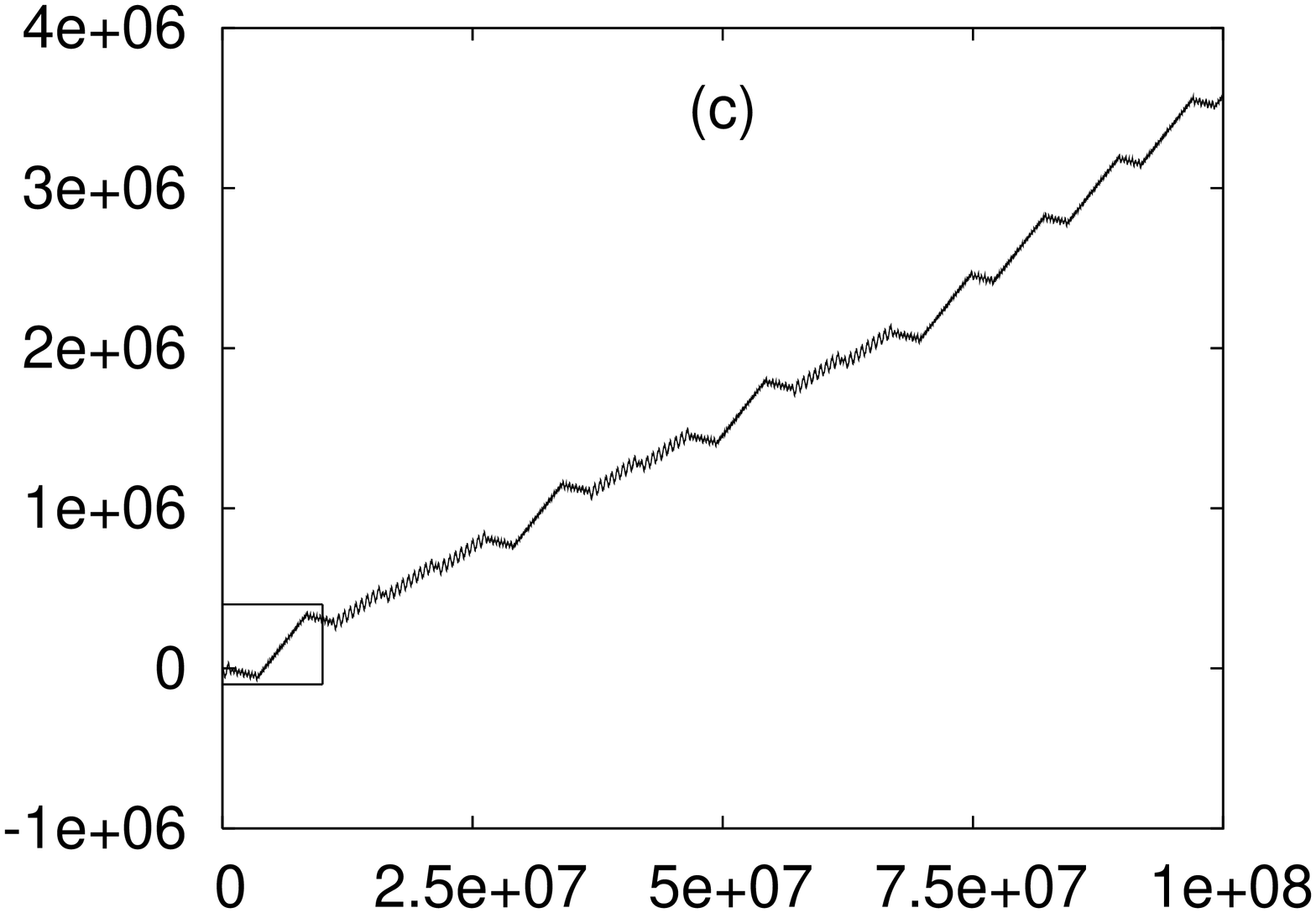 , angle=-90 , width=7cm }
  \epsfig{file=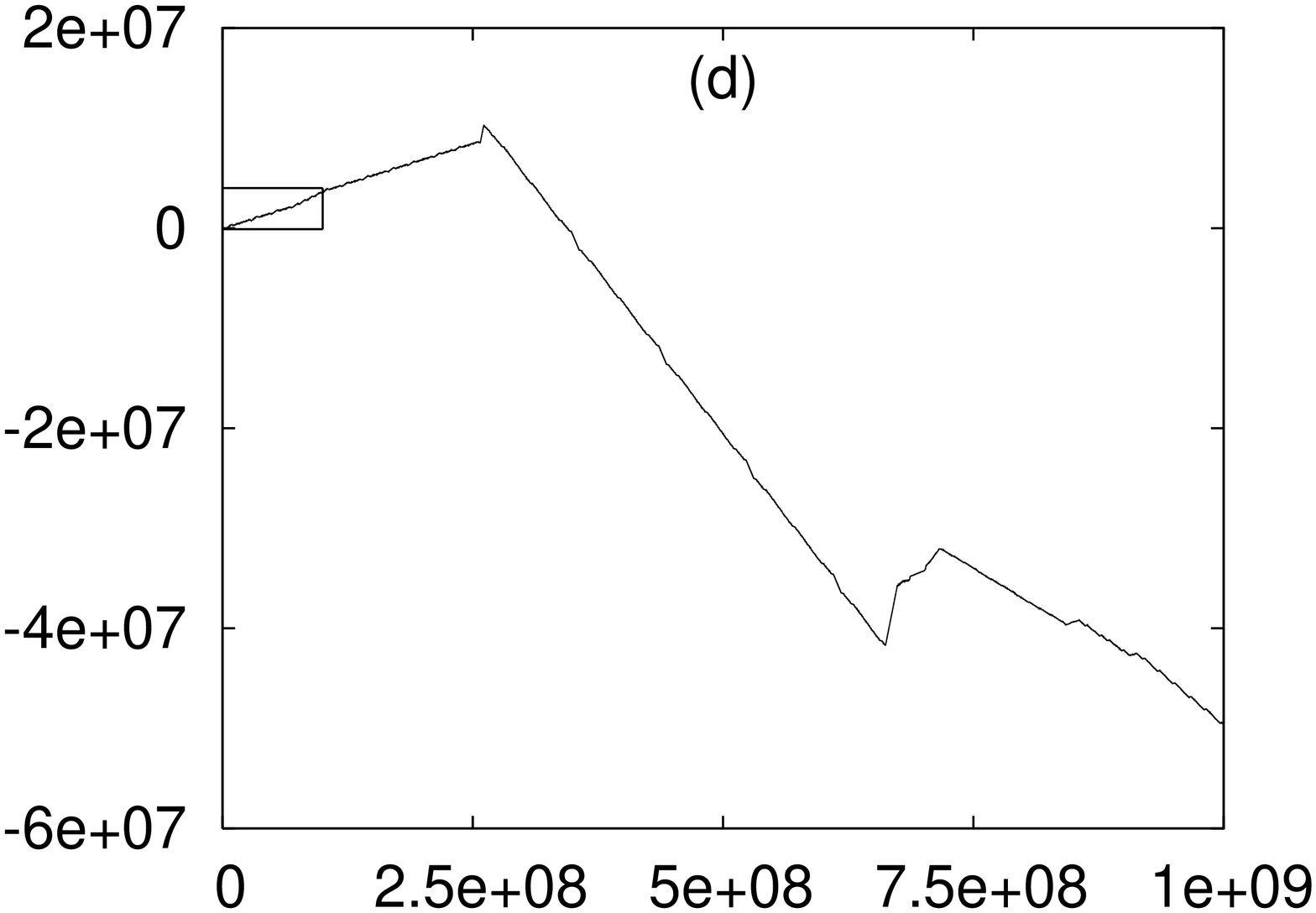 , angle=-90 , width=7cm }
  \caption{Example of the time evolution of a particle displacement for the $\Delta y/\Delta x=0.25$
  system ($d=0.5\Delta y$), for a sequence of four different time-scales. Boxes indicate the position of the
  previous graph in the sequence.}
  \label{par-s2}
\end{figure}

\begin{figure}
  \centering
  \epsfig{file=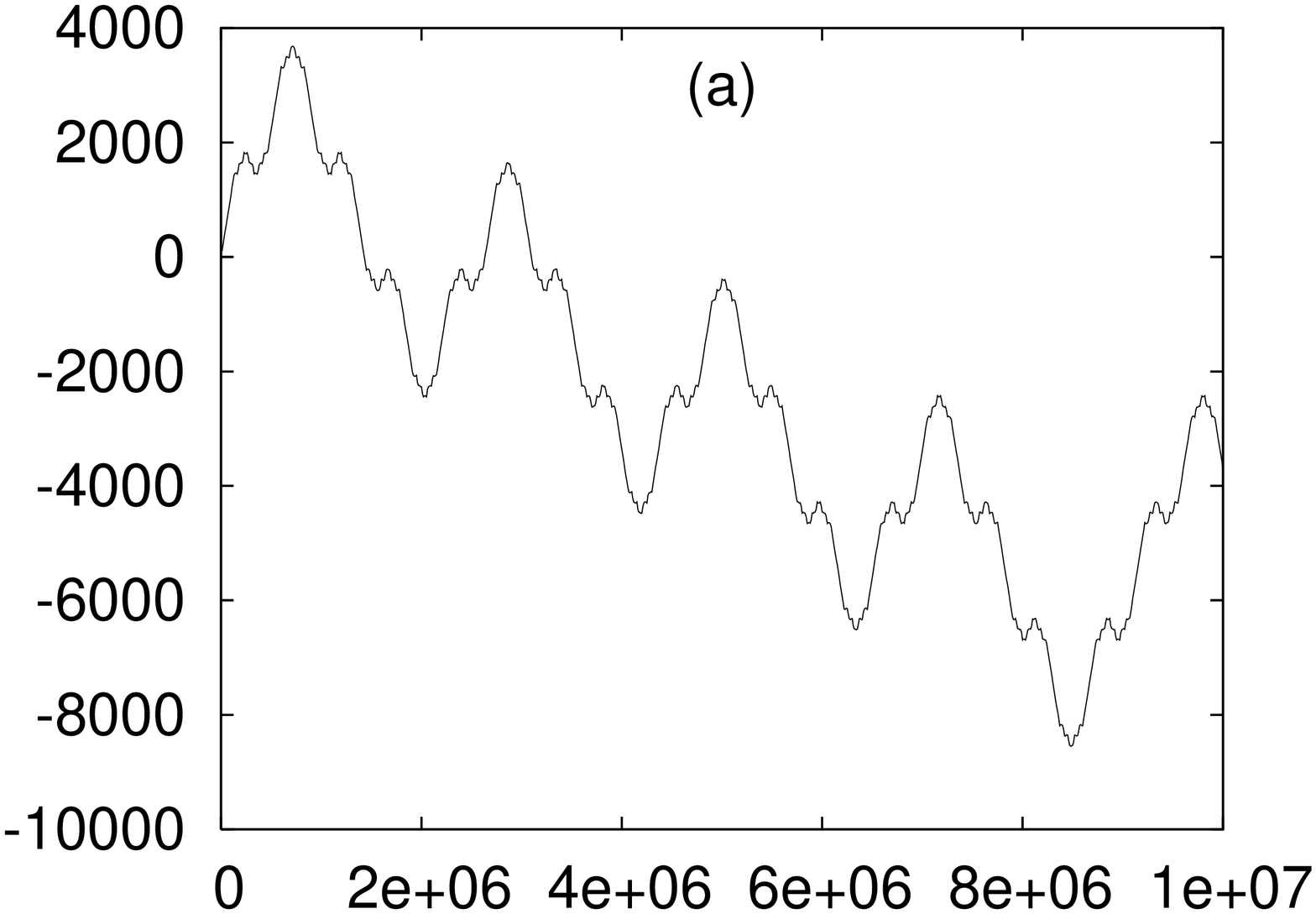 , angle=-90 ,  width=7cm }
  \epsfig{file=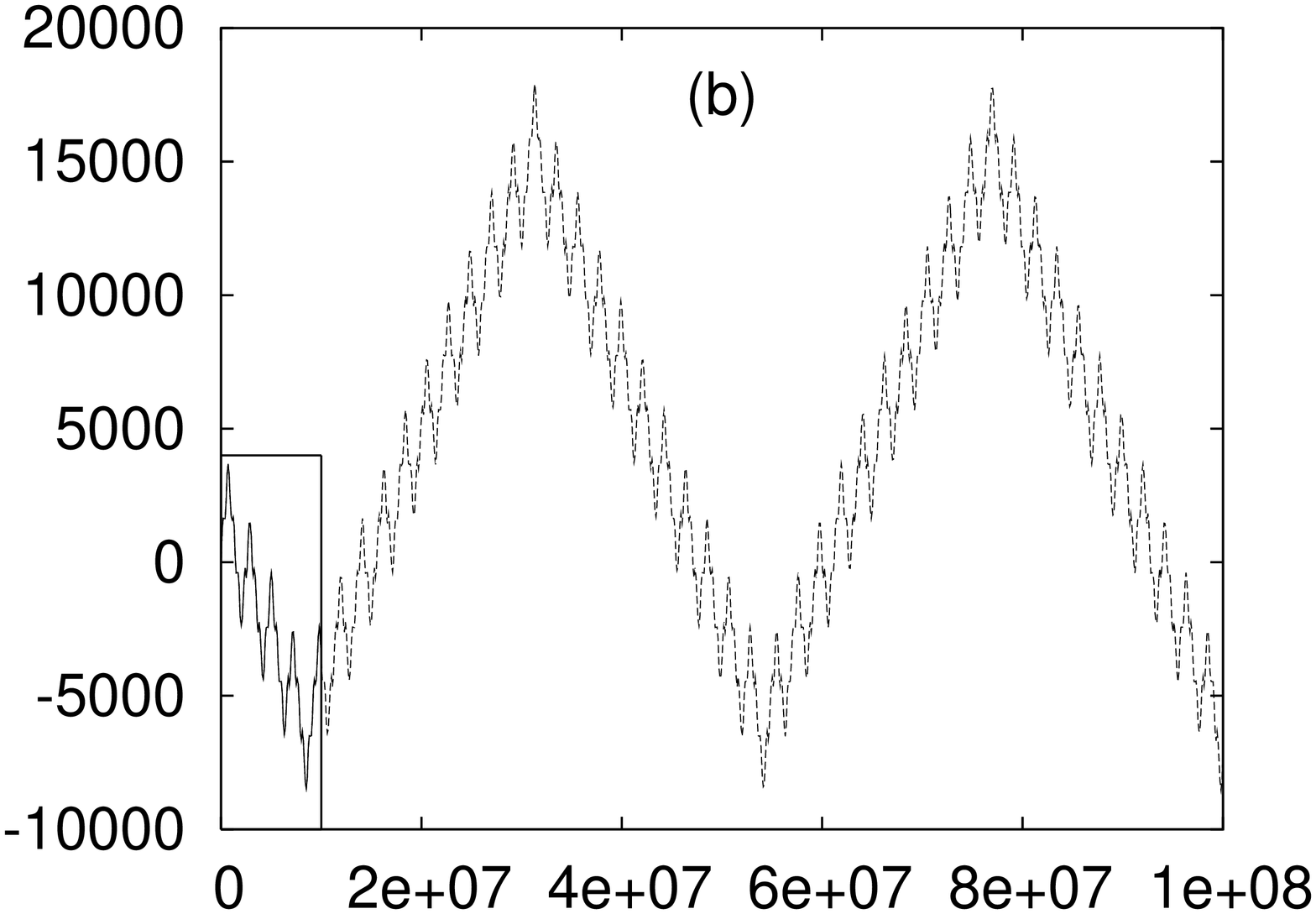 , angle=-90 ,  width=7cm }
  \epsfig{file=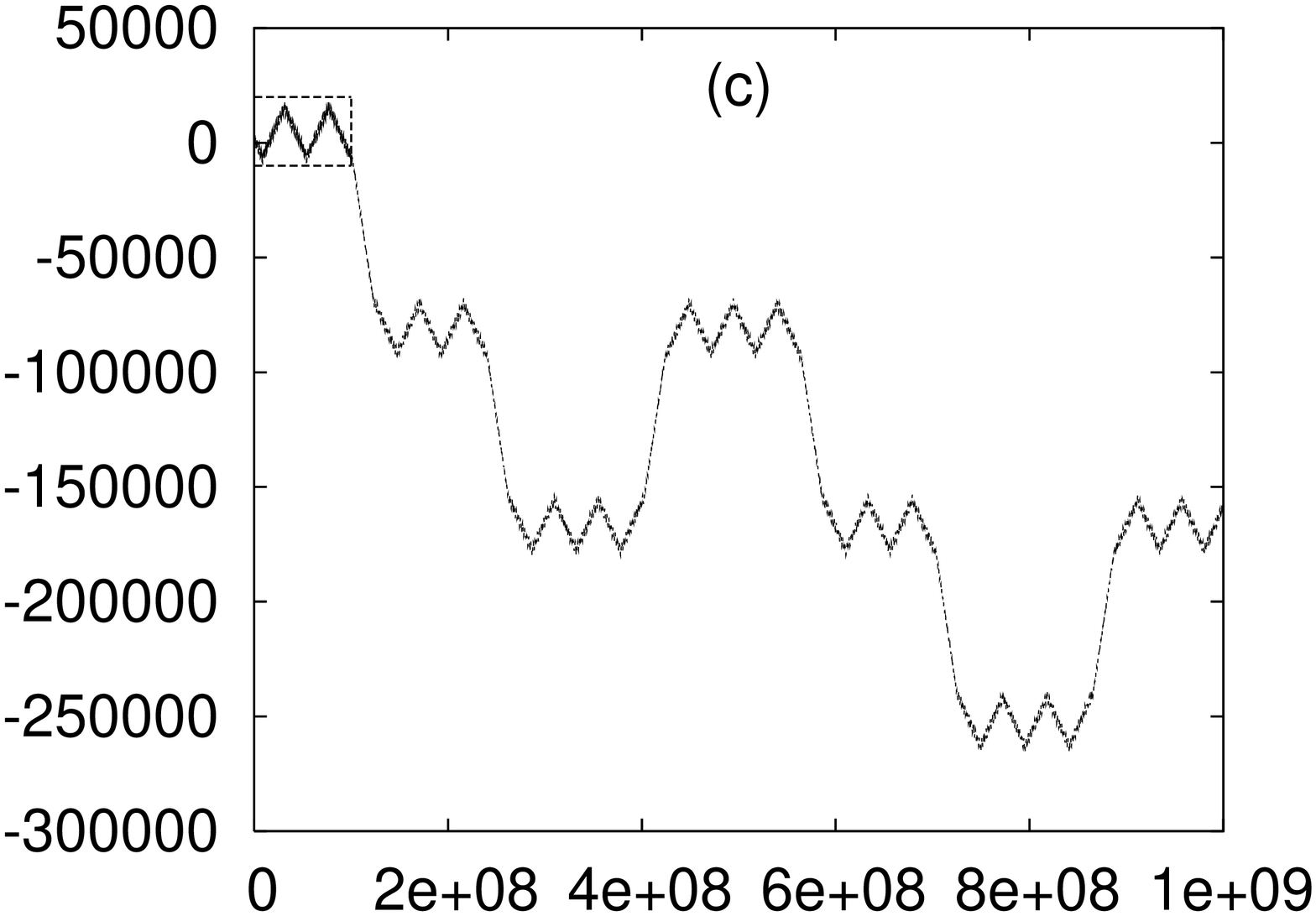 , angle=-90 ,  width=7cm }
  \epsfig{file=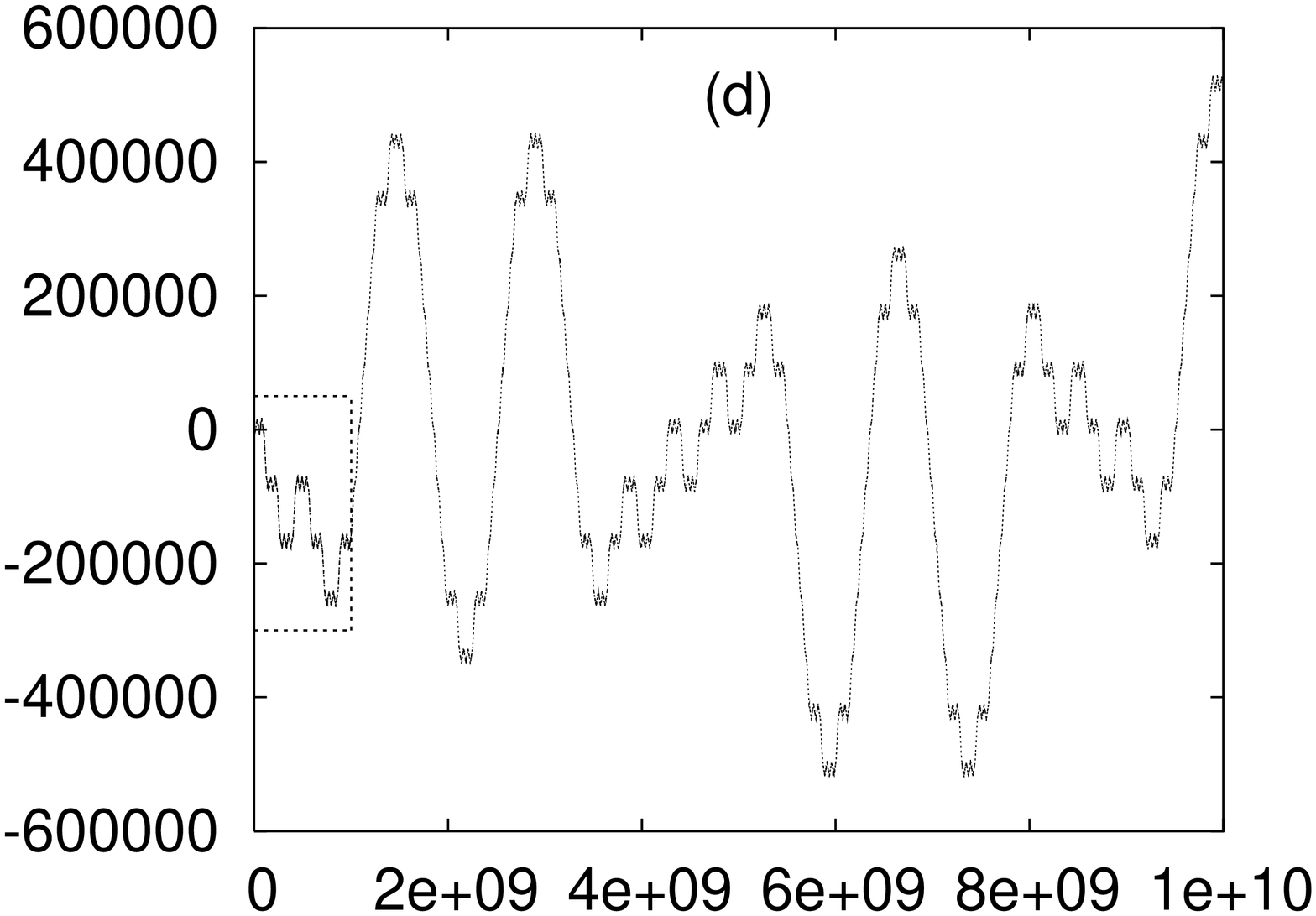 , angle=-90 ,  width=7cm }
  \caption{Example of the time evolution of a particle displacement for the $\Delta y/\Delta x=1$
  system ($d=2\Delta y$), for a sequence of four different time-scales. Boxes indicate the position of the
  previous graph in the sequence.}
  \label{par-s3}
\end{figure}

Finally, we turn our attention to the behaviour of the displacements $s_x(t)$ for individual initial
conditions, as a function of time, which we expect to give further insight into the manner in which
the momentum space is explored. In \fig{par-s2}, we show $s_x(t)$ as a function of time over four
different time-scales for a single particle trajectory in a system where $\Delta y/\Delta
x=0.25$. Again, such a trajectory is typical of the results observed for the parallel irrational
polygonal billiards examined. In \fig{par-s3}, we show an analogous set of results for transport of
a particle in a (rational) $\theta = \pi/4$ system.

In \fig{par-s2}a, the dynamics appear somewhat random on the scale of $10^6$ time units, although on closer 
inspection it is clear that certain sections of the trajectory are repeated. Indeed, the occurrence of 
repeated segments is more obvious in \fig{par-s2}b, where after an initial transience of $10^6$ steps, the 
simulation appears to reach a periodic orbit, before changing to another orbit in the interval at about 
3.5$\times10^6$, then reverting to the original orbit at around 8$\times10^6$ time units. On a longer time 
scale (\fig{par-s2}c), the transport appears to alternate between these two almost periodic orbits. However, 
on the scale of $10^9$ time units (\fig{par-s2}d), the evolution of $s_x$ takes on quite a
different appearance, bearing a similarity with what one might observe for the
random motion of a particle in a low-density gas.
Such a resemblance suggests that, on this time-scale, there could be sufficient
memory loss of the preceding momentum values for the trajectory to appear random. Even at
this stage, however, the number of distinct momenta visited by the particle is still
limited to the order to 10-100. 

In contrast, we recall that the transport behaviour in the $\theta=\pi/4$ system is restricted to a maximum 
of four distinct momenta, for each initial condition. Despite this apparently strong limitation, a richness 
in behaviour is still evident. In \fig{par-s3}, we observe strongly recurrent behaviour on all time-scales 
observed, although the nature of the recurrence varies on all observed time-scales, and is likely to 
continue to do so at larger and larger scales. We note that 
\fig{par-s2}d and \fig{par-s3}c represent the evolution of $s_x$ over the same time-length --- however, 
while this evolution appears random in the irrational system, the motion is regular in the rational 
case. The regularity is less trivial in \fig{par-s3}d, with four small steps between 
the first two large steps, but only three small steps between the second two large steps, showing that the 
apparent regularity doesn't make the dynamics easily predictable. 

\begin{figure}
  \centering
  \epsfig{file=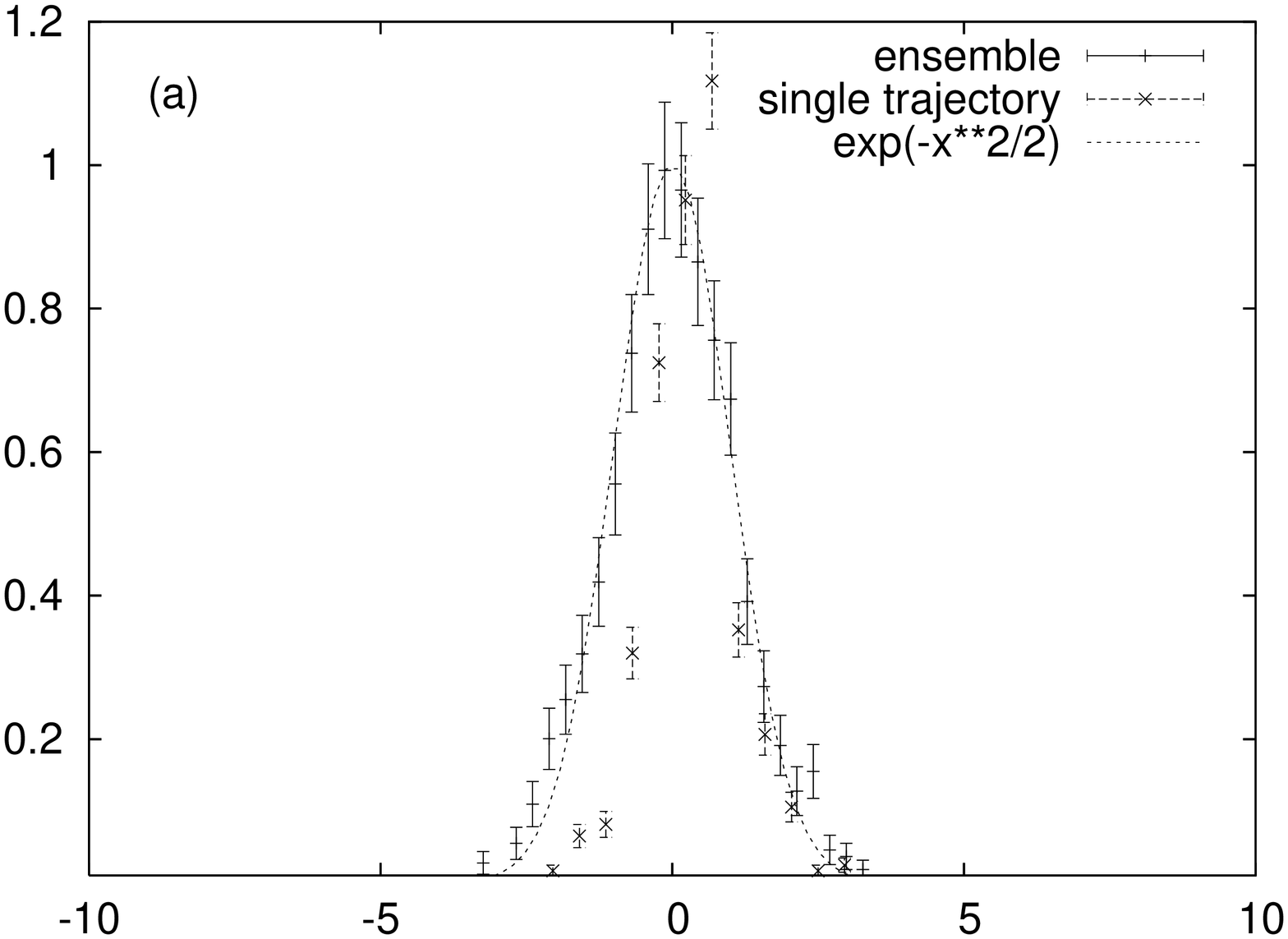 , angle=-90 , width=7cm }
  \epsfig{file=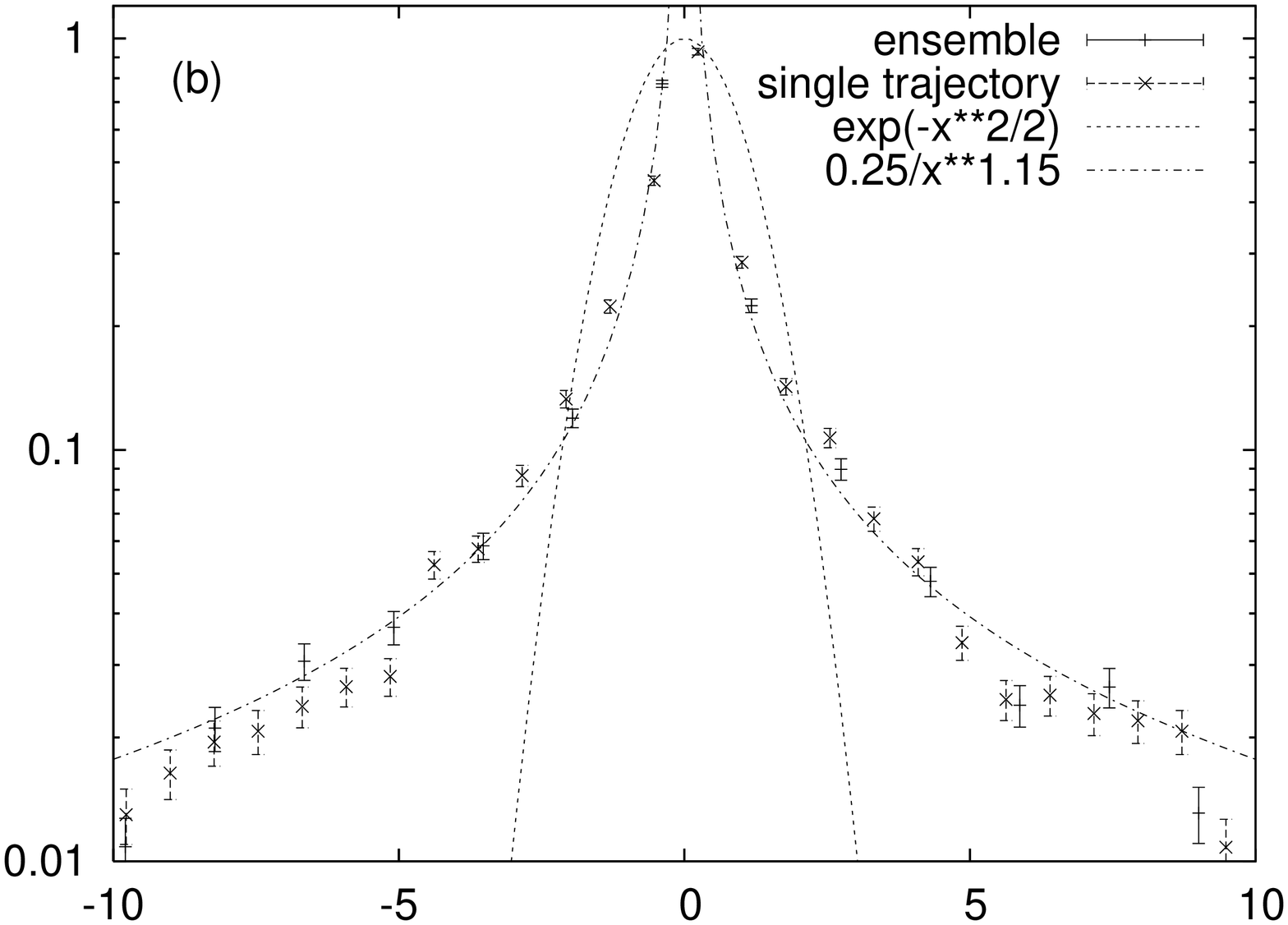 , angle=-90 , width=7cm }
  \caption{Histogram of displacements collected along a single trajectory, and from the ensemble of trajectories, for the (a) $\Delta y/\Delta x=0.25$ ($d=0.5\Delta y$), and (b) $\Delta y/\Delta x=1$ ($d=2.0\Delta y$) systems.}
  \label{par-s2-hist}
\end{figure}

In \fig{par-s2-hist}a, we show the distribution of displacements obtained by dividing the $10^9$ time-unit trajectory for the $\Delta y/\Delta x=0.25$ system into segments of shorter time periods, in order to compare with the approximation to the ensemble distribution obtained from averaging over trajectories with independent initial conditions. It is clear from the figure that the two distributions are significantly different --- while the ensemble distribution demonstrates the near-Gaussian properties observed earlier, the distribution from the single trajectory is significantly skewed (as could be expected from \fig{par-s2}), and has a distribution that is much narrower than that of the ensemble. The correlations observed along the trajectory in \fig{par-s2} lead to a distribution of displacements that is not at all characteristic of the ensemble, up to a time of one billion time units.

In \fig{par-s2-hist}b, we show the distribution of displacements obtained by dividing the $10^{10}$ time-unit trajectory for the $\Delta y/\Delta x=1$ system into segments, as done above. In constrast with the results for the $\Delta y/\Delta x=0.25$ system, the distribution obtained in this fashion shows excellent agreement with ensemble distribution of trajectories. 

\vskip 5pt
\noi
{\bf Remark 2. }{\it The overall transport is ultimately \emph{slower} in the rational systems than 
  in the irrational systems. Furthermore, despite the limitations on the velocities that each single 
  trajectory can take, which differ from trajectory to trajectory, the distributions of displacements
  exhibited by one rational trajectory equals the distribution of the ensemble, while this is not
  the case for the irrational trajectories\,!}

\subsubsection{Unparallel walls, collective behaviours}
\label{unpar-coll-eq}

\begin{figure}
  \centering
  \epsfig{file=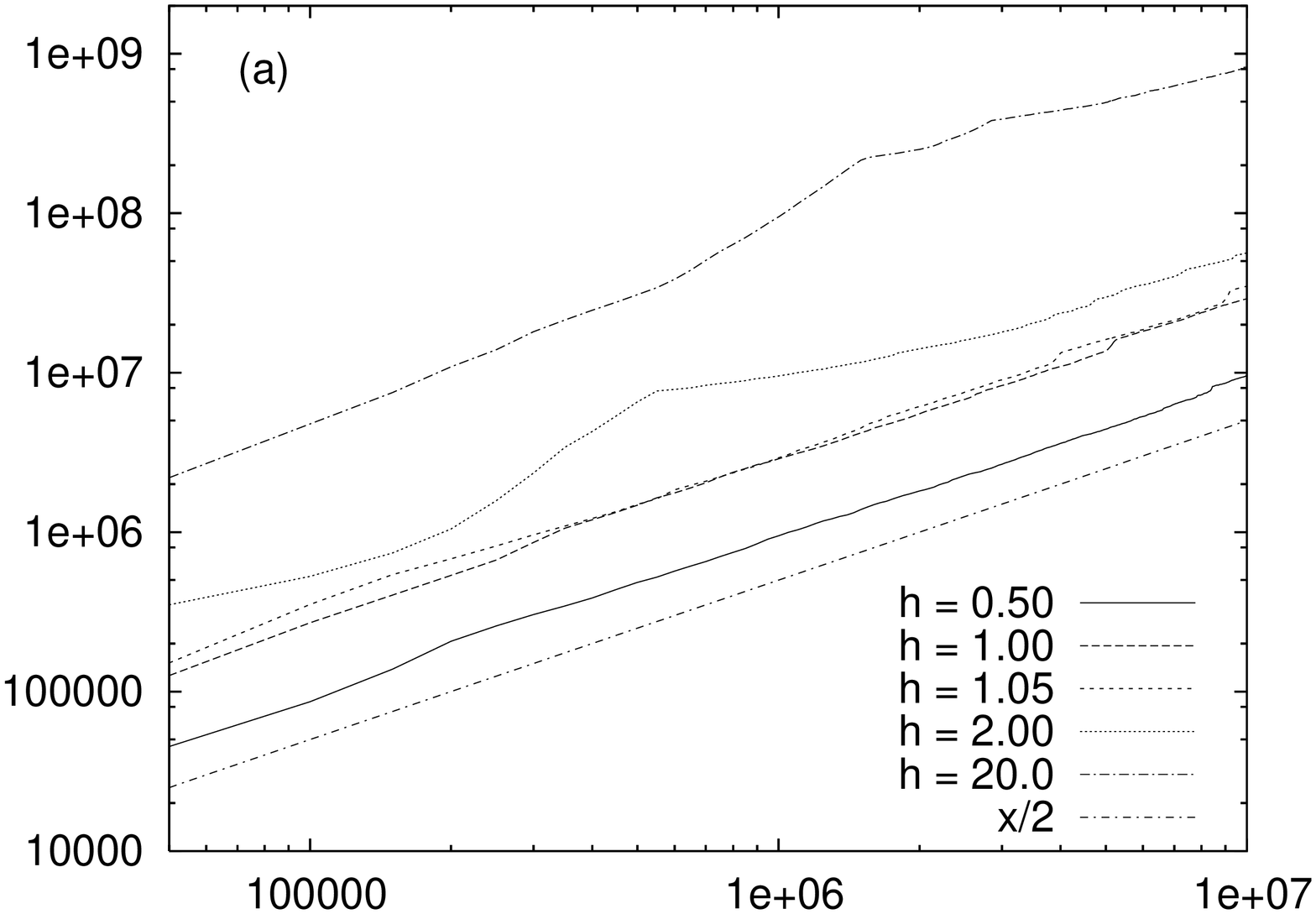 , angle=-90 , width=8cm }
  \epsfig{file=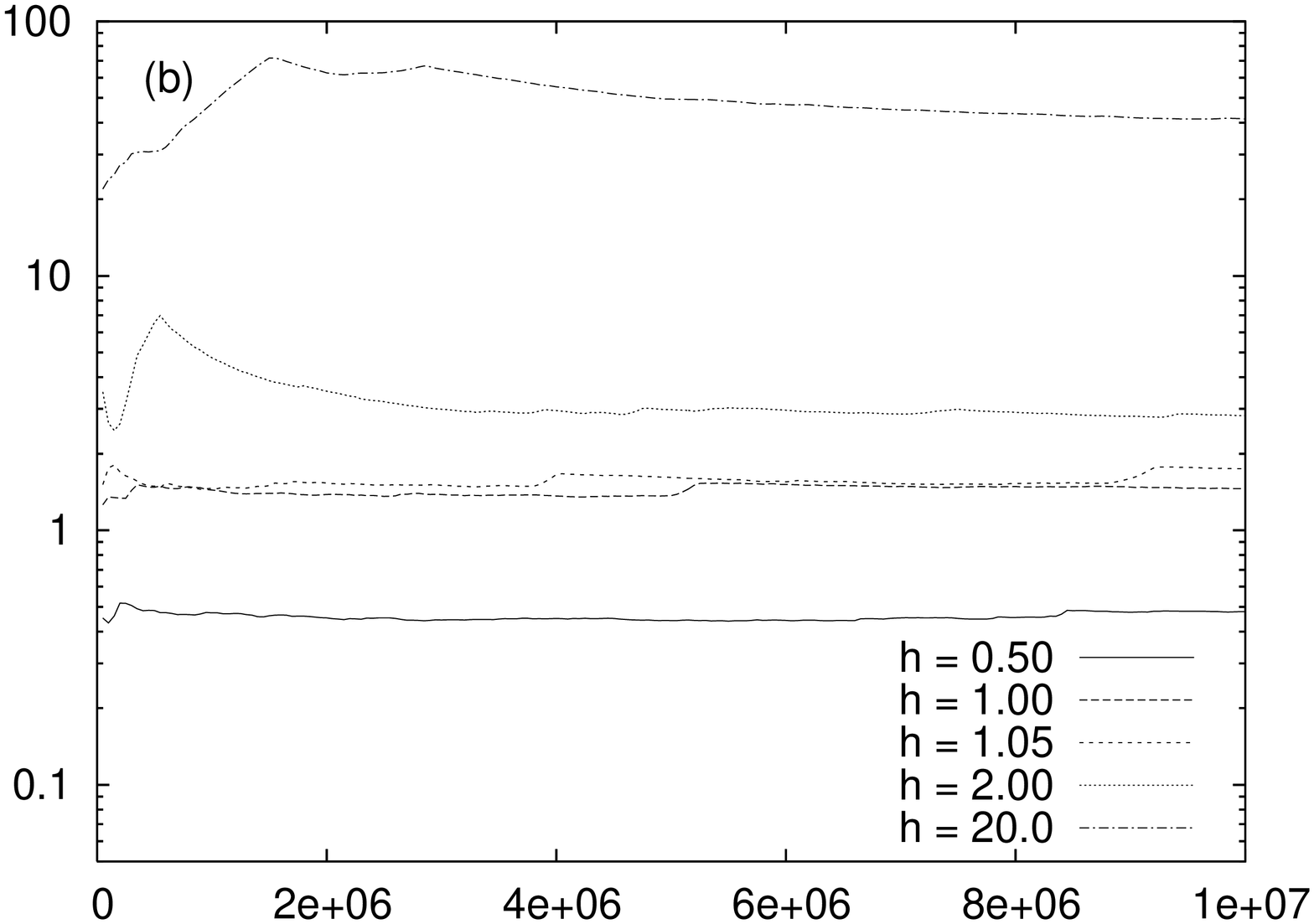 , angle=-90 , width=8cm }
  \caption{Evolution of (a) the mean-squared displacement, and (b) the diffusion coefficient, 
           for transport in systems with unparallel saw-tooth walls --- 
           $\Delta y_t/\Delta x = 0.62$, $\Delta y_b/\Delta x = 0.65, d=0.5d_c,d_c,1.05d_c,2d_c$ and $20d_c$.}
  \label{unpar-msd}
\end{figure}

We have examined the transport properties of a series of unparallel saw-tooth systems, chosen such 
that the ratios $\Delta y_b/\Delta x$ lie in the vicinity of the golden ratio. In \fig{unpar-msd} 
we show the behaviour of the mean-square displacement, and the estimated diffusion coefficient, as 
a function of time, for a series of unparallel saw-tooth systems where
$\Delta x = 0.5, \Delta y_t/\Delta x = 0.62$, and $\Delta y_b/\Delta x = 0.65$. 
In analogy to the results above, we have examined these systems at the same range of pore heights 
based on the mean interior pore height $d_c=(\Delta y_t+\Delta y_b)/2$ at which the horizon becomes 
infinite --- at $d=0.5d_c,d_c,1.05d_c,2d_c$ and $20d_c$.

Despite containing data from $10^4$ independent initial conditions, the data in \fig{unpar-msd} exhibit features 
suggesting that they have not yet converged to a final result. These features correspond to significant jumps 
in the mean-squared displacements (and consequently the finite-time estimate of the diffusion coefficient), 
resulting from short bursts of quasi-periodic behaviour (i.e.\ ballistic transport). We note from \fig{unpar-msd} 
that the effects of the bursts appears to grow as the horizon is opened. Similar jumps are observed in 
the time-evolution of the super Burnett coefficients, indicating that the bursts contribute to driving the
system away from a Gaussian distribution.

\begin{figure}
  \centering
  \epsfig{file=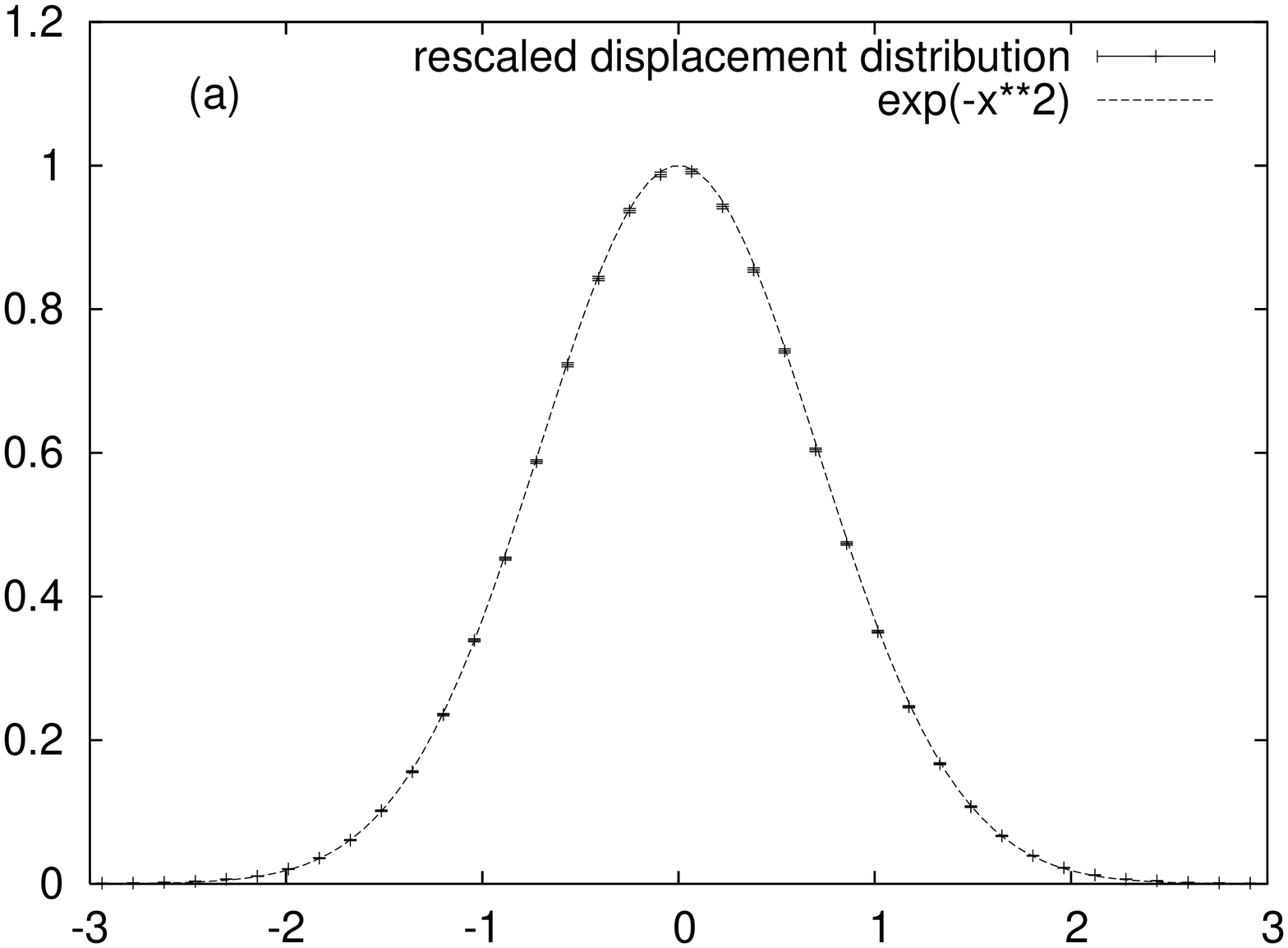 , angle=-90 , width=7cm }
  \epsfig{file=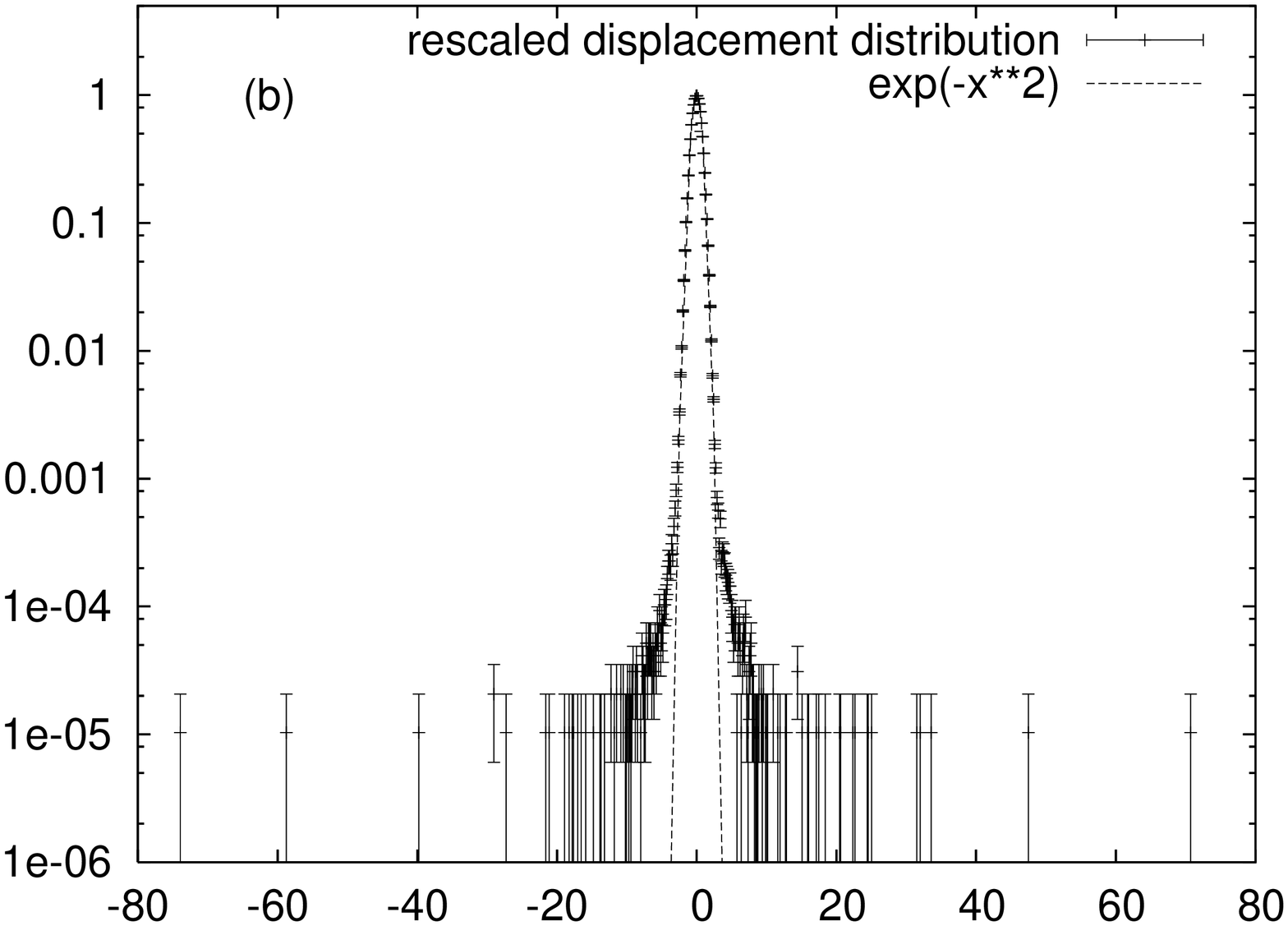 , angle=-90 , width=7cm }
  \caption{Displacements over intervals of $10^5$ time units, combining contributions from all $10^4$ initial
  conditions, for transport in the system $\Delta y_t/\Delta x = 0.62$, $\Delta y_b/\Delta x = 0.65$, for $d=0.5d_c,d_c,1.05d_c,2d_c$
  and $20d_c$: (a) within three standard deviations; and (b) over the full range of displacements (shown using a
  log-linear plot).}
  \label{unpar-displacements}
\end{figure}

To give some sense of the size and frequency of these bursts, we show the distribution of displacements 
obtained over intervals of $10^5$ time units, combining contributions from all $10^4$ initial conditions, 
in \fig{unpar-displacements}. We observe excellent agreement with the Gaussian distribution, out to several 
standard deviations. However, in the tail of the distribution we find a non-negligible contribution from 
large-scale displacements, to which we attribute the behaviour of the super Burnett coefficients. These 
contributions correspond to the bursts observed in \fig{unpar-msd}, and it is clear that a huge number of 
initial conditions would be required before the overall effect of this tail distribution could be realised 
by simulation.

\tabl{unpar-flat} shows the values of the exponents obtained from fitting the data from the observed systems 
to \q{defn:gamma}, and obtain behaviour that is close to diffusive. Errors in the Table were determined using
a Marquardt-Levenberg non-linear least squares fit. The significant errors in the data arise from the ballistic
bursts, so that a least-squares error estimate, more appropriate for random errors, may not be as appropriate 
here. However, the least-squares error estimates still provide useful information regarding the relative errors 
in the data obtained for the various systems.

From \fig{unpar-displacements} it is clear that the bursts only occur for a small number of particles. We have 
found that, in each case, only 2 or 3 initial conditions are responsible for the `significant' bursts --- that is 
to say, if the contribution from these 2 or 3 particles (in 10000) is neglected, the resultant behaviour is 
diffusive within statistical error, and the fluctuations all lie within error about this mean diffusive behaviour.
Within statistical error, one could conjecture that the effect of these `significant' bursts is not sufficient to drive the
behaviour away from diffusive behaviour, given the decay back to diffusive transport observed in \fig{unpar-msd}
after each burst. Clearly, however, it cannot be excluded that the effect of these bursts is to drive the transport 
at a rate somewhat faster than diffusive, either with an exponent slightly greater than 1, or with some slower 
correction, such as the $\ln t$ correction for the Sinai billiard. We note, with the Sinai billiard in mind, that 
the bursts responsible for this potentially super-diffusive behaviour are observed in systems with both open and
closed horizons.

\begin{table}
  \centering
  \begin{tabular}{c|c|c|c|c|c|c|}
    $\Delta y_t/\Delta x$ & $\Delta y_b/\Delta x$ & \multicolumn{5}{c|}{saw-tooth systems}  \\
                 &              & $0.5 \Delta y$  & $1.0 \Delta y$  & $1.05 \Delta y$ & $2.0 \Delta y$  & $20 \Delta y$  \\
\hline
        0.62     &    0.63      &     1.00(2)     &      1.02(2)    &     0.97(3)     &     1.03(7)     &     0.72(3)    \\
        0.62     &    0.64      &     1.00(1)     &      1.2(1)     &     1.03(3)     &     1.19(7)     &     1.10(5)    \\
        0.62     &    0.65      &     0.99(2)     &      1.02(2)    &     1.02(3)     &     0.97(6)     &     1.13(5)    \\
  \end{tabular}
\caption{Equilibrium transport exponents: For unparallel saw-tooth boundary base triangles with height-width ratio
$\Delta y/\Delta x$. For each (mean) pore height tested, the observed exponent out to $10^6$ time units (of the 
order of $10^6$--$10^7$ collisions) is given. The number of initial conditions used to compute averages 
varies from 2000 up to $10^4$. Numbers in brackets correspond to error estimates from Marquardt-Levenberg 
least-squares fits.}
  \label{unpar-flat}
\end{table}

\begin{figure}
  \centering
  \epsfig{file=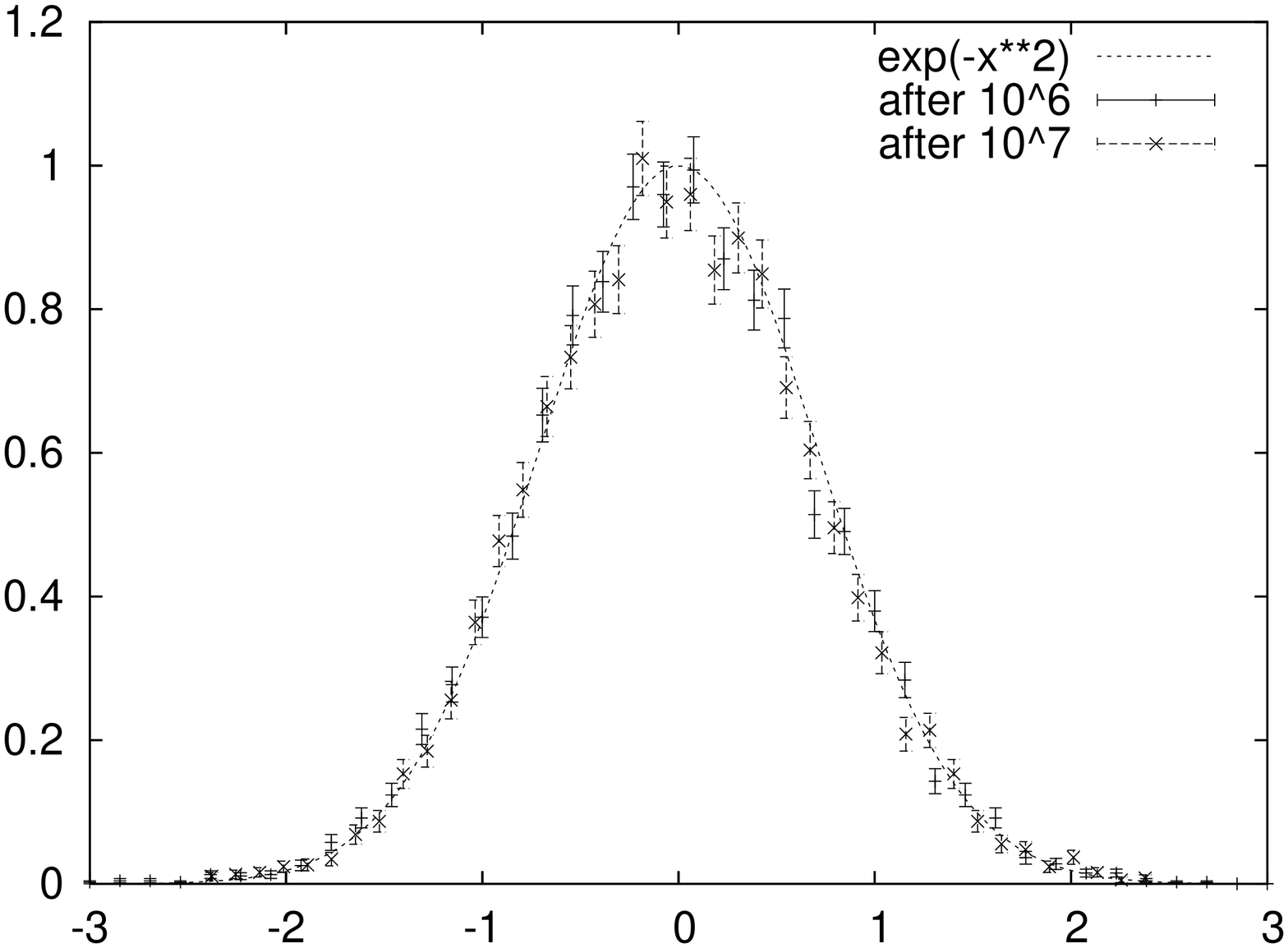 , angle=-90 , width=12cm }
  \caption{Distribution of final displacements for unparallel saw-tooth systems where $\Delta y_t/\Delta x = 0.62, \Delta y_b/\Delta x = 0.65, d=0.5d_c$.}
  \label{unpar-displace}
\end{figure}

In \fig{unpar-displace}, we show the distribution of displacements after $10^6$ time units, and
after $10^7$ time units, from the trajectories of individual particles, again noting that the
initial distribution is effectively a delta function since all particles begin from the same unit
cell at $s_x=0$. Even after $10^6$ time units, the distribution of displacements is very well fit by
a Gaussian, consistent with our observations of a diffusive transport rate, and in contrast to the
results for the parallel walls.

As with the case of parallel walls, the momenta do not appear to be significantly correlated over the course of the 
simulation, and there is stronger evidence in this case of a convergence to a uniform distribution than in the
parallel case.

\subsubsection{Unparallel walls, individual behaviours}

As with the parallel systems, we have considered the behaviour along longer individual trajectories,
to compare with the ensemble behaviour. For unparallel walls, we find that there is strong agreement
between the individual and ensemble behaviours, in terms of the distribution of both the momentum
orientations and of the displacements.
Sequences of momenta appear random along a trajectory, and the distribution of displacements
obtained along a single trajectory demonstrates an excellent Gaussian fit, in agreement with
the ensemble results.


\subsection{Nonequilibrium}
\label{nonequilibrium}

An alternative method of studying the transport behaviour is to consider the dynamics in the
presence of an external field. For thermodynamic fluids, we expect that the transport coefficient
determined in the linear response regime (i.e.\ in the zero-field limit) for the nonequilibrium
fluid corresponds to that obtained from the equilibrium fluid properties. It is not clear, \emph{a
  priori}, that such an equivalence holds for the system examined in this paper, so we examine the
nonequilibrium transport properties, as a dynamical system of interest in its own right, as well as
to compare its behaviour with that of the equilibrium counterpart.

\subsubsection{Parallel}
\label{parallel}
\begin{figure}
  \centering
  \epsfig{file=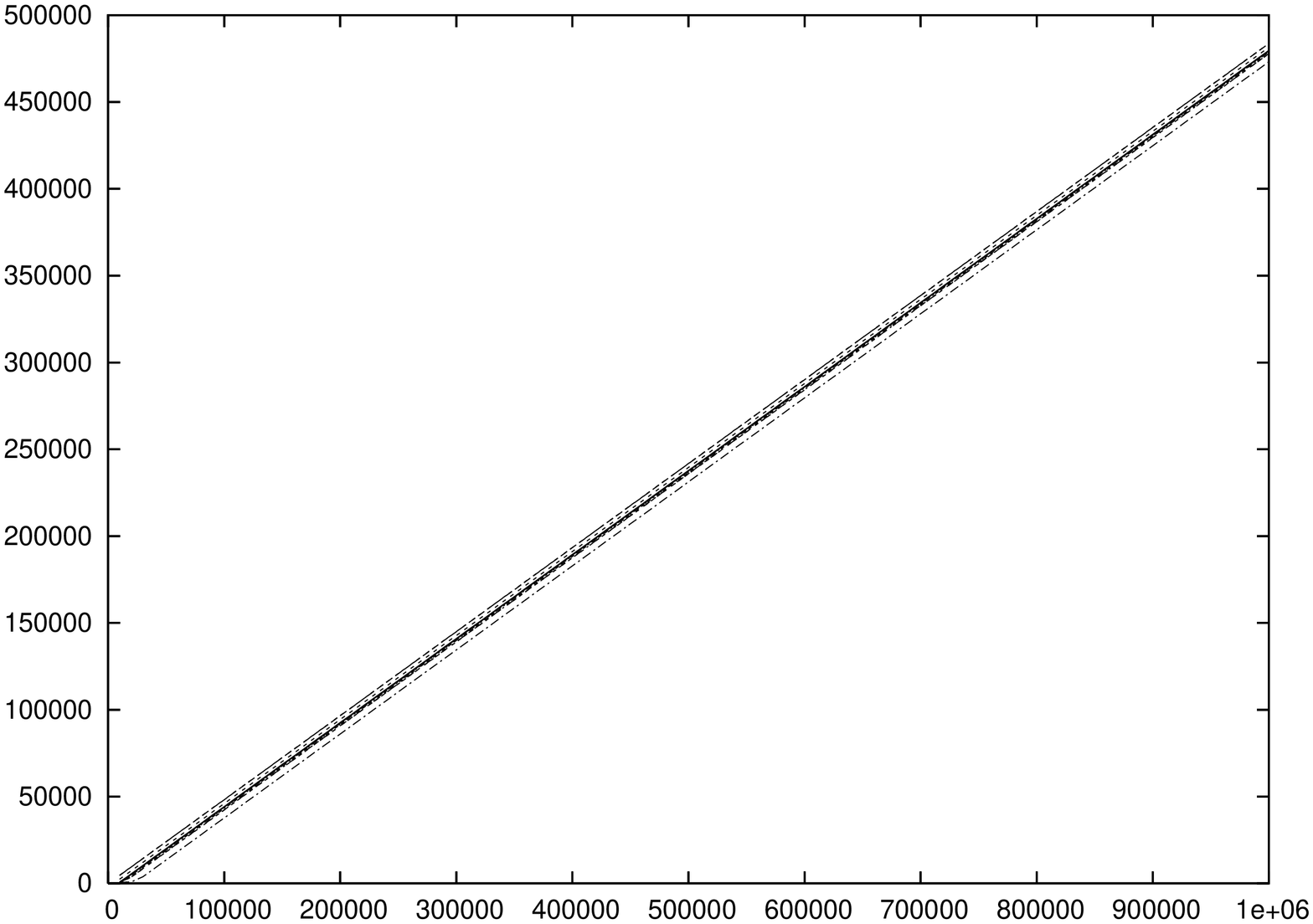 , angle=-90 , width=8cm }
  \epsfig{file=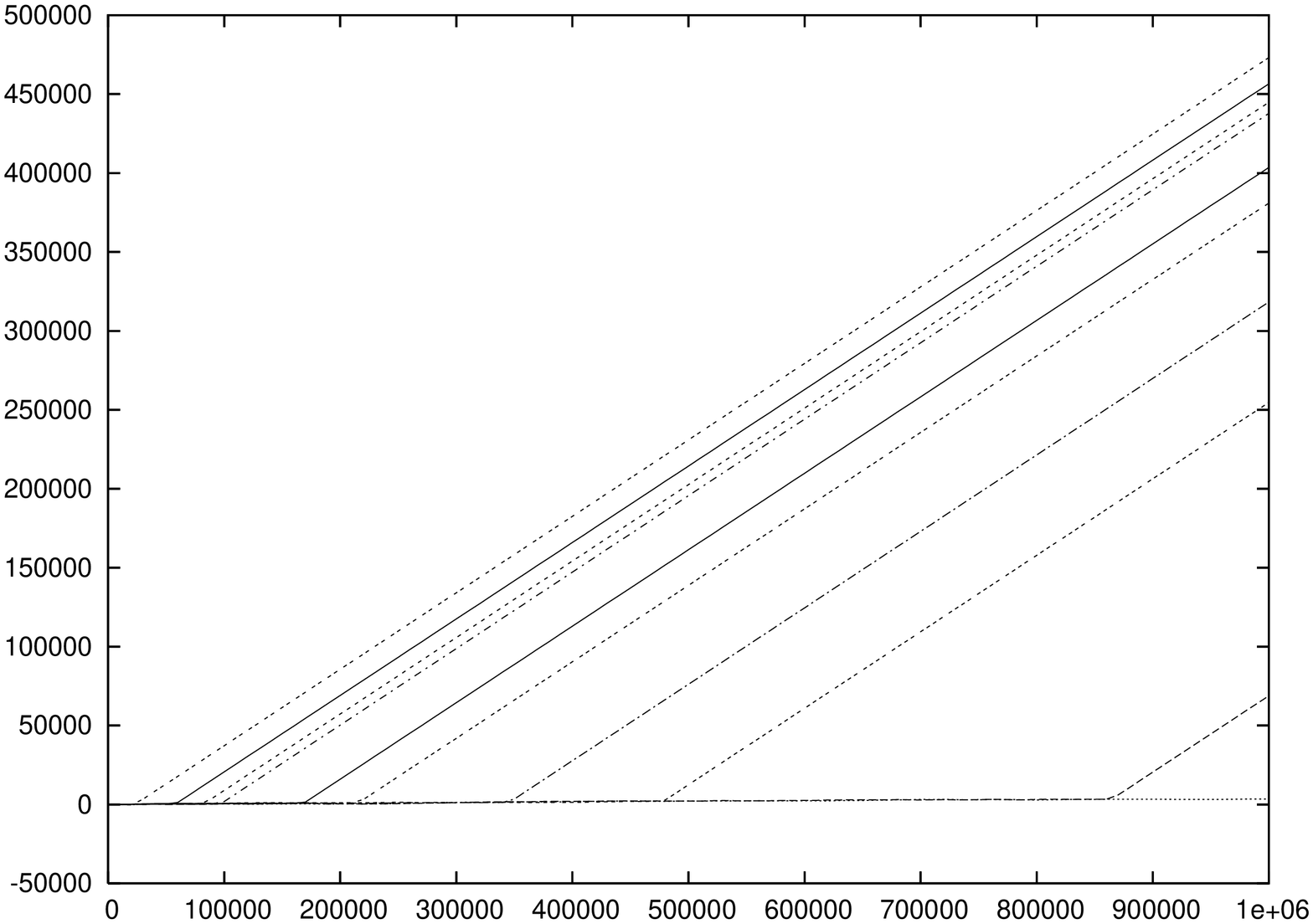 , angle=-90 , width=8cm }
  \epsfig{file=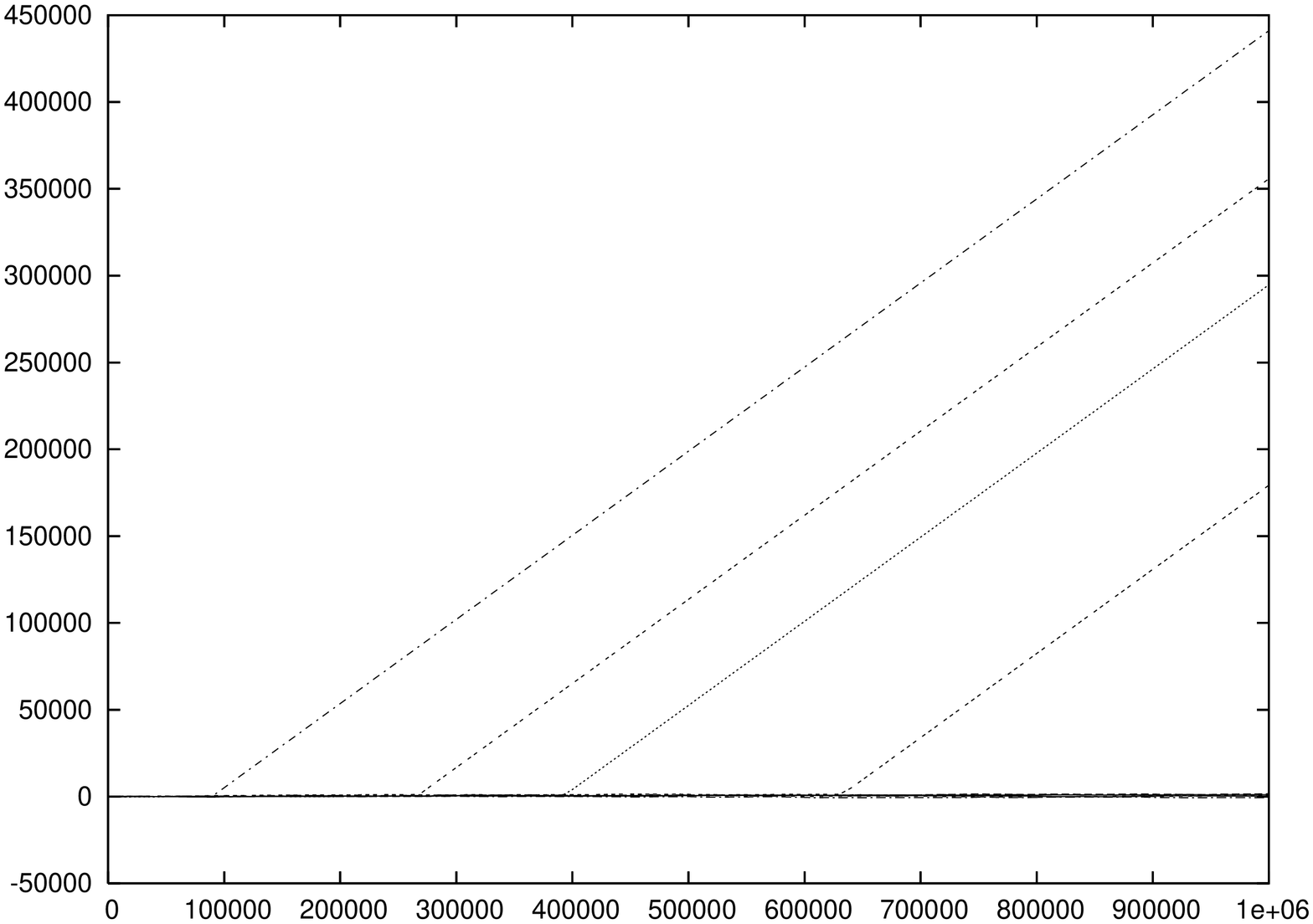 , angle=-90 , width=8cm }
  \epsfig{file=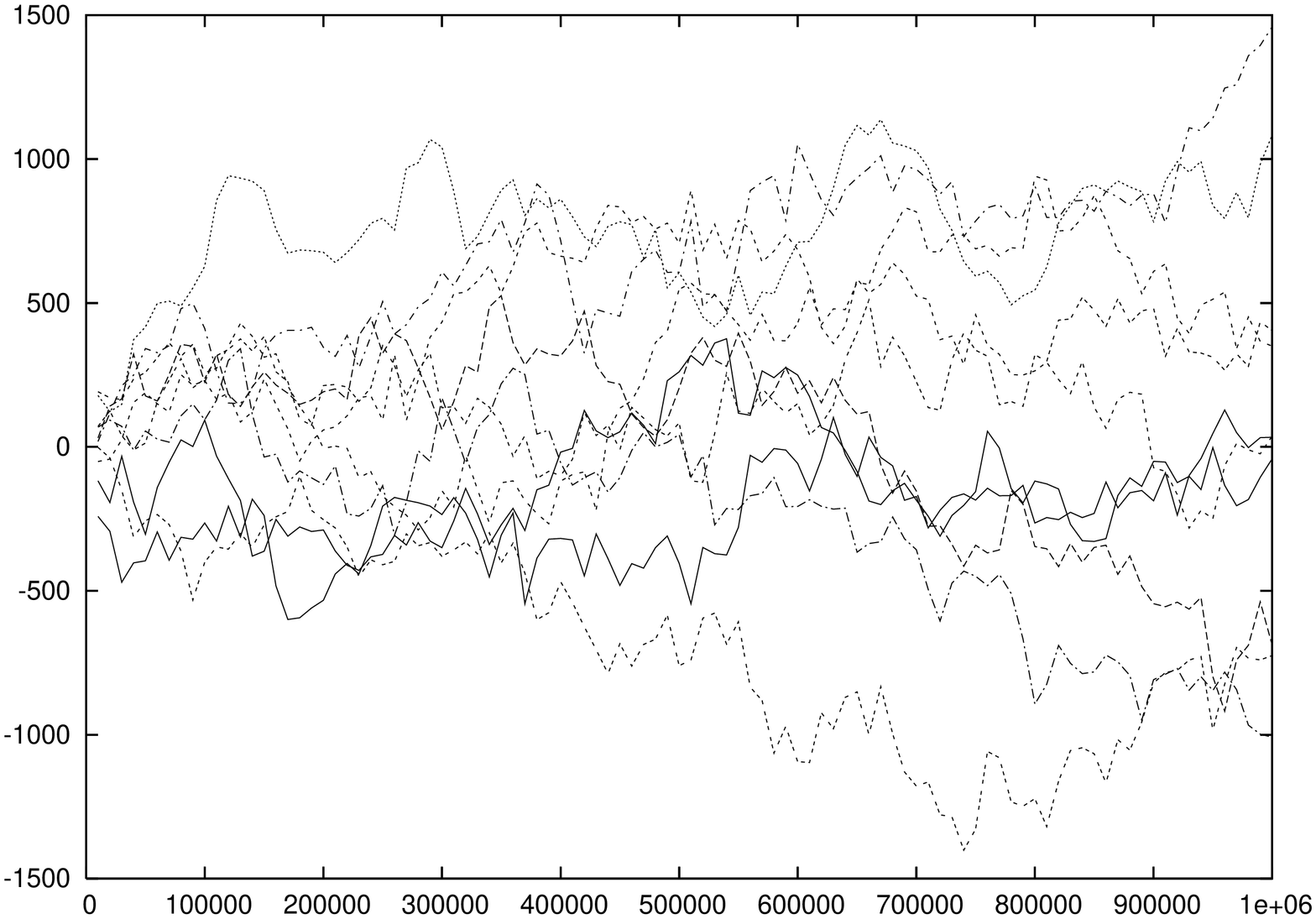 , angle=-90 , width=8cm }
  \caption{Displacement versus time for ten trajectories for the parallel $\Delta y/\Delta x = 2$ system, 
  for (a) $\epsilon=0.1$, (b) $\epsilon=0.01$, (c) $\epsilon=10^{-3}$, and (d) $\epsilon=10^{-4}$.}
  \label{par-200-traj}
\end{figure}

\begin{figure}
  \centering
  \epsfig{file=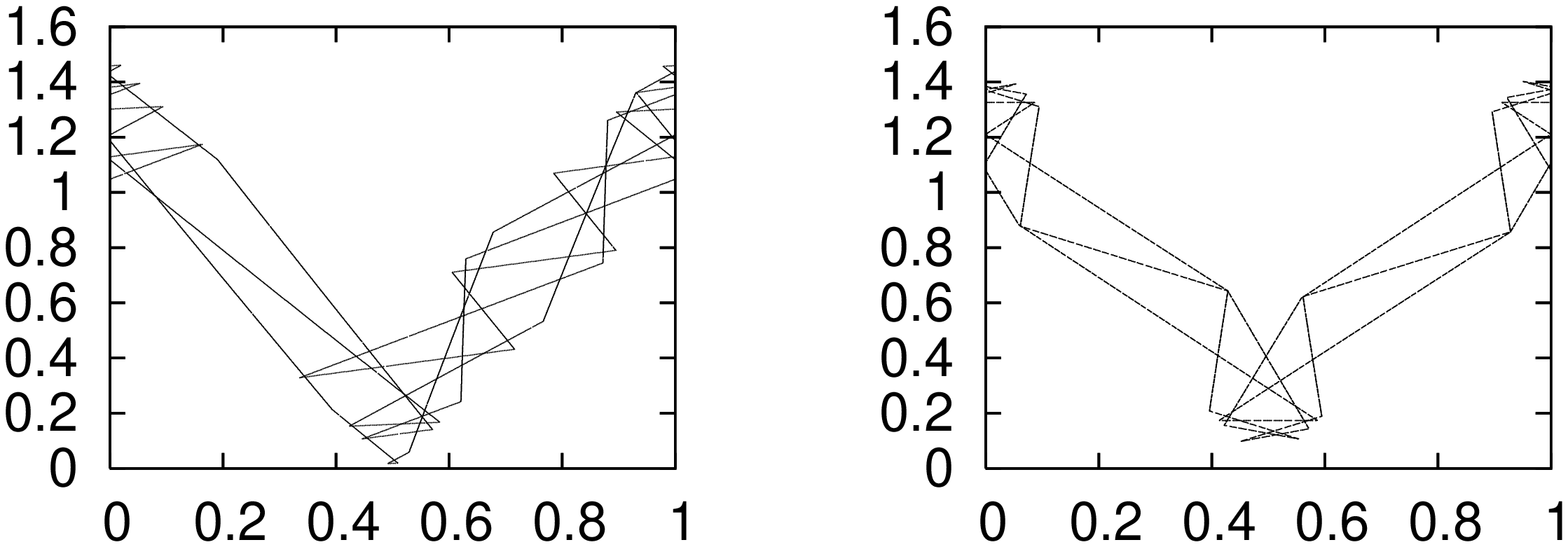, bb = 300 60 555 760, angle=-90 , width=12cm } 
  \caption{Two distinct periodic trajectories of the finite field transport for the parallel $\Delta y/\Delta x = 2$ system 
  for $\epsilon=0.01$. The left hand side orbit has period $\tau = 12.393386...$ time units with 39 reflections per orbit. The
  right hand side orbit has period $\tau = 9.6950889...$ time units with 33 reflections per orbit.}
  \label{par-200-per}
\end{figure}

For the non-zero-field estimates of the transport coefficient for the $\Delta y=2$ system, with 
$d = 0.5 \Delta y$, for the four different field strenghts $\epsilon=0.1, 0.01, 10^{-3}, 10^{-4}$, we find that
the qualitative behaviour of the particles is highly, and unpredictably, dependent on the field strength. At the
highest field, $\epsilon=0.1$, (\fig{par-200-traj}a) all trajectories have similar qualitative and quantitative
properties, demonstrating a ``fluid-like'' response to the external field --- particles are driven in the direction
of the field, and the fluctuations about a mean transport rate are small, since the field is strong.
At $\epsilon=0.01$, however, the trajectories exhibit two distinct transport phases --- an initial phase where the
particle trajectories fluctuate about a mean motion due to the driving field, and a second ballistic phase, where
the trajectory finds a periodic orbit (\fig{par-200-traj}b). At this field, two distinct orbits were noted --- one
consisting of 33 reflections, with period $\tau =  9.6950889...$ and mean net speed $v_b\approx0.31$, the other 
consisting of 39 reflections, with period $\tau = 12.393386...$ and mean net speed of $v_b\approx0.24$. The 
orbits are shown in \fig{par-200-per}. In particular, we note the existence of distinct periodic orbits to 
which the different trajectories converge, demonstrating that the dynamics at this field strength is 
\emph{not ergodic}. At lower fields, a transition to periodic orbits is much rarer --- however, bursts of 
almost-periodic orbits are observed, which decay after relatively short times to revert to the previous 
apparently random behaviour (\fig{par-200-traj}d).

This transition from apparently random transport to ballistic transport has been observed
previously in a similar nonequilibrium system with straight walls \cite{jsp099_0857}. There,
a transition time was observed between these two transport behaviours, which varied
as the inverse square of the external field strength. Our results appear to be consistent
with such a relationship, inasmuch as the fraction of observed periodic orbits (indicating
ballistic transport) decreases as the field strength increases. We expect that longer
simulations would produce a larger fraction of such periodic orbits, and suggest that the reason
we do not observe any ballistic transport in the weakest field is due to the much larger
time-scale on which such a transition would take place. Unlike the systems in \cite{jsp099_0857}, 
where only the ensemble behaviour was studied, and a smooth transition from diffusive to ballistic 
transport was observed at finite times, here we consider the single trajectories, and find that their
individual transitions are sharp and occur apparently at times not limited by any upper bound.

We show the estimates of $D(\epsilon)$ for the applied fields $\epsilon$ for the parallel $\Delta
y/\Delta x = 2$ system in Column 2 of \tabl{zero-field-sum}. Over such a range of fields, we would
expect to observe either convergence to well-defined value (the diffusion coefficient), or divergent
behaviour (bounded by the thermostat). However, we note that $D(\epsilon)$ does not diverge in the
zero-field limit, as we would expect for a real fluid which exhibited superdiffusive
equilibrium behaviour (such as plug flow).

\begin{table}
  \centering
  \begin{tabular}{c|c|c||c|c|c}
    field strength $\epsilon$ &    parallel system    &    parallel system    &   unparallel system   & unparallel system   \\
                              &         (raw)         &      (corrected)      &         (raw)         &    (corrected)      \\
\hline
         0.1                  &  $0.0935 \pm 0.0001$  &  $0.0935 \pm 0.0001$  &  $ 2.390 \pm 0.003$   &  $2.390 \pm 0.003$  \\  
         0.01                 &  $9.75   \pm 0.2$     &  $0.205  \pm 0.008$   &  $18.3   \pm 0.2$     &  $0.24  \pm 0.04$   \\  
         0.001                &  $0.57   \pm 0.04$    &  $0.57   \pm 0.04$    &  $11     \pm 1$       &  $0.19  \pm 0.02$   \\  
         0.0001               &  $0.28   \pm 0.4$     &  $0.28   \pm 0.4$     &  $ 1.2   \pm 1$       &  $0.28  \pm 0.14$   \\
\hline
       equilibrium            &       $\infty$        &        $\infty$       &  $0.386  \pm 0.05$    &  $0.386  \pm 0.05$
  \end{tabular}
  \caption{Best estimates of the finite field diffusion coefficient $D(\epsilon)$ for $0.1<\epsilon<10^{-4}$ for the parallel $\Delta y/\Delta x = 2$ system, and the unparallel $\Delta y_t/\Delta x = 0.62, \Delta y_b/\Delta x = 0.65, d=d_c$ system, showing the raw data and data corrected for periodic orbits. Also included, for comparison, are the equilibrium results for the parallel case (exhibiting super-diffusive behaviour) and the unparallel case. Data was collected over $10^6$ time units.}
  \label{zero-field-sum}
\end{table}

\subsubsection{Non-parallel}
\label{non-parallel}

In contrast to the results for the parallel systems, we find that the individual particle
trajectories for the $\Delta y_t/\Delta x = 0.62, \Delta y_b/\Delta x = 0.65, d = 0.5 d_c$
system display the same characteristic behaviour for the four different field strengths
$\epsilon=0.1, 0.01, 10^{-3}, 10^{-4}$ examined. Trajectories are typified by an initial apparently
random transient behaviour, followed by a transition to ballistic transport in a periodic orbit. 
The transition time also appears to vary
inversely with the field strength. As a consequence, the fraction of trajectories observed to undergo
a transition within $10^6$ time units decreases with decreasing field, and therefore the lifetime of
the non-ballisitic regime grows.

\begin{figure}
  \centering
  \epsfig{file=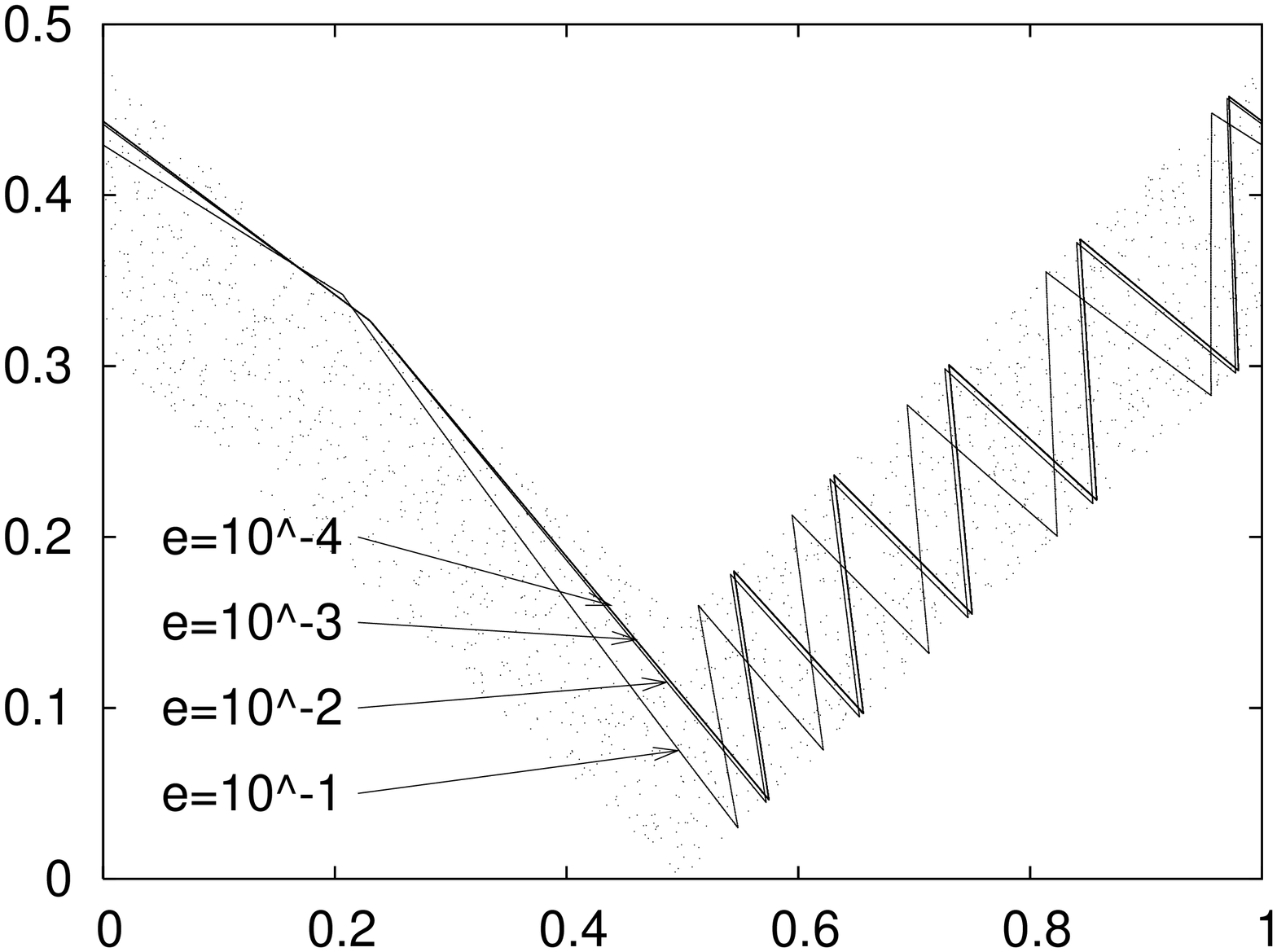 , angle=-90 , width=12cm }
  \caption{(Attractive) periodic trajectories of the finite field transport for the $\Delta y_t/\Delta x = 0.62, \Delta y_b/\Delta x = 0.65, d=d_c$ system, for $0.1<\epsilon<10^{-4}$. The trajectories for the two weakest fields are indistinguishable on the scale of the figure. Cloud shown for aesthetic reasons.}
  \label{unpar-065-per}
\end{figure}

For a given field strength $\epsilon$, each ballistic
trajectory appears to have the same mean net speed. For each field strength considered, this mean
net speed appears to be $v_b\approx0.48$. We find that the individual trajectories converge to a
single periodic orbit that is independent of the initial conditions, but dependent on the field
strength. These orbits are shown in \fig{unpar-065-per}. We note that these periodic orbits appear
to be instances of a continuous family of periodic orbits, converging to a limiting orbit in the
zero-field limit. However, this limiting orbit is never observed in the equilibrium trajectories,
where it is no longer an attractor. 

The existence of a family of attractive periodic orbits, converging to a precise periodic orbit in the
zero-field limit, precludes the existence of a linear regime, and is significantly at odds with what
one would expect of an ergodic, diffusive system in nonequilibrium. For such systems, we would
anticipate the convergence of the nonequilibrium finite-field estimates $D(\epsilon)$ to the
equilibrium value $D$ in the zero-field limit. However, the limiting periodic orbit with a
finite net speed $v_b\approx0.48$ implies that the finite-field estimate $D(\epsilon)$ must
\emph{diverge} in the zero-field limit.

However, as noted above, the
fraction of trajectories that turn ballistic decreases with decreasing field, so that in the $\ze \to 0$
limit we expect this fraction to tend to 0. Furthermore, the behaviour of the trajectories \emph{before}
they reach a periodic orbit
remains apparently random, and responsive to the field. Consequently, we consider the following
approach to constructing the linear regime. At any given time, we neglect
those trajectories already captured by a periodic orbit (ie whose transport has become ballistic), and
we define a finite diffusion coefficient based upon the remaining trajectories. Usually, one
defines the nonequilibrium diffusion coefficient (following the Green-Kubo result \q{defn:D-noneq})
by taking the limits $t\rightarrow0$ and $\epsilon\rightarrow0$ separately --- here, this is no
longer possible if we wish to avoid periodic trajectories, so the limits must be taken
simultaneously. 

Therefore, given a sufficiently large ensemble of $N$ particles, assume that the fraction $\nu(\ze;t)$ of particles
which are still in the diffusive regime, at time $t$ and for field strength $\ze$, is well-defined.
Let $M(N;\ze;t)$ be the subset of indices in
$\{1,\ldots,N\}$ of these still-diffusive trajectories. Assume that 
for every $\zd > 0$ there is a field $\ze_\zd > 0$ such that $\nu(\ze;t)
\ge 1 - \zd$ if $\ze < \ze_\zd$ and $N$ is sufficiently large, as seems to be the case in our simulations. For simplicity, assume that there is a function $\ze=\ze(t;\zd)$,
such that $\nu(\ze(t;\zd);t) = 1 - \zd/2$. Finally, 
define $N(\zd)=[2l/\zd]$, where $[\cdot]$ represents the integer part, and choose integer $l$ so large
that $N(\zd)$ is sufficiently large for the fraction  $\zd/2$ to be sufficiently finely realized, for any $\zd>0$.

\vskip 5pt

\noi
{\bf Definition 4. }{\it For every $\zd > 0$, distribute at random (with respect 
to the Lebesgue measure) $N(\zd)$ initial conditions in the phase space. For each
of them and for fixed $t$, consider
$$
D_i(\ze;t) = \frac{k T v_{i,x}}{m \ze} ~, \quad
i\in\{1,\ldots,N\} ~,
$$
where $v_{i,x}$ is the mean x-component of the velocity of the $i$-th trajectory 
(and other variables defined as in \q{defn:D-eps}). The {\em nonequilibrium estimate} 
of the diffusion coefficient, if it exists, is defined by
\be
\begin{split}
D_{\text{ne}} 
&= \lim_{\zd \to 0} \lim_{t \to \infty}
\frac{1}{N(\zd) \nu(\ze(t;\zd);t)} \sum_{i\in M(N(\zd);\ze;t)} D_i(\ze(t;\delta);t) \\
&= \lim_{\zd \to 0} \lim_{t \to \infty}
\ \frac{kT}{m\ze(t;\delta)} \ \frac{1}{N(\zd)\nu(\ze(t;\zd);t)} \ \frac{1}{t} \sum_{i\in M(N(\zd);\ze;t)} s_{xi}(t),
\end{split}
\ee
} \noi
Thus, for a given choice of $\zd$, and $t$, we evaluate the estimate of the diffusion
coefficient (\q{defn:D-eps}) considering only those trajectories which have not been captured by a periodic orbit.
Then, the limit $t \to \infty$ includes also the limit $\ze \to 0$, and the
limit $\zd \to 0$ includes the limit $N \to \infty$. We can now define the following:

\vskip 5pt
\noi
{\bf Definition 5. }{\it
If the equilibrium system has a finite diffusion coefficient $D$, and $D_{\text{ne}}=D$, the system is said to 
have a {\em linear regime}.
}

In \tabl{zero-field-sum} we report the `corrected' estimates $D(\epsilon)$ for the various fields,
where we neglect contributions from ballistic trajectories. In the third column we report estimates
for the parallel $\Delta y/\Delta x = 2$ system, and in the fifth column we report estimates for the
unparallel  $\Delta y_t/\Delta x = 0.62, \Delta y_b/\Delta x = 0.65, d = 0.5 d_c$ system. We note
that the `corrected' data for the parallel $\Delta y/\Delta x = 2$ system are consistent with a
divergent trend (taking into account the large statistical error in the weakest field). For the unparallel
system, the equilibrium and nonequilibrium transport properties become more consistent, although
they are not conclusively so, because of the large error bars produced by the statistical analysis
at low field. Such noise is typical of NEMD simulations at low field, where the signal-to-noise
ratio becomes low. Typical NEMD simulations, however, would still converge to yield the same transport
coefficient in the long-time limit --- in our non-chaotic systems, the transition from random to ballistic 
behaviour effectively places an upper bound on the time during which random behaviour can be observed. 
The estimated errors in the diffusion coefficient obtained will therefore depend on the rate at which
the mean transition time increases with decreasing field.

\vskip 5pt
\noi
{\bf Remark 3. }{\it This subsection leads us to conclude that the use of thermostats needs some form of
chaos, whether permanent or transient, in order for a linear regime to be observed. This is a cause for concern 
in the case of thermostatted {\em non-interacting} particle systems.}

\section{Transport Complexity}
\label{complexity}

The properties of the dynamics in these systems are strongly dependent on the values of $\theta_t$
and $\theta_b$, and their relative properties --- whether they divide one another, and whether they
divide $\pi$. This sensitivity to the boundary conditions is strongly ``unthermodynamic'' in nature,
and owes its origin
to the absence of inter-particle interactions. In real molecular dynamical
systems, chaotic behaviour acts as a ``double-edged sword'' --- on the one hand, the sensitivity to
initial conditions ensures that trajectories diverge exponentially; on the other hand, under such dynamics
almost all trajectories explore the phase space in
such a way that they display the same macroscopic behaviour. 
In the current systems, the dynamics appears to display a sensitivity to initial
conditions despite the absence of dynamical chaos. However, the dynamics is not always sufficiently
mixing to cause all trajectories to demonstrate the same average behaviour. By a judicious choice of
saw-tooth angles, one can produce dynamics which appear consistent with macroscopic notions of
diffusion. Clearly, however, in
the absense of intermolecular interactions, one can only expect to describe those aspects of
thermodynamics which are based on purely convective processes. Such models lack the essential
element of momentum or energy exchange that would be required to describe processes that involve
conduction, such as heat transfer or viscous effects, and consequently one cannot draw conclusions on
such processes based on the systems of noninteracting particles, such as those examined here.

Similar considerations appear to apply to chaotic noninteracting particle systems, as they also 
have been found to have thermodynamic properties irregularly dependent on the geometry of the 
medium \cite{LlNiRoMo95,HKG02,ZK04,Kla10}. However, in these systems, the presence of chaos induces
a higher stability, which is manifested in the fact that the transport coefficient, but not the 
transport \emph{law}, is affected by the geometry. The transport coefficient, in particular, turns out
to be a close-to-continuous function 
\footnote{while the function may not be continuous, it can be represented by a function $D(\xi)+\epsilon(\xi)$ 
 for geometry parameter $\xi$, where $\max |\epsilon(\xi)|$ is of the same (if not smaller) order than the 
 errors in measurement, or  where $\max |\epsilon(\xi)|/(\partial D/\partial \xi)$ is very small compared 
 to the range of $\xi$ of interest, or both.}
, which may, or may not, be a fractal curve, depending on the case. As conceptually interesting
as this question is, it does not bear serious consequences for the predicted transport,
when the parameters determining the geometry are only known with a finite precision. Indeed, this
implies a small error in the transport coefficient, which results in a small uncertainty
on the transported quantities. In other words, transport is ``predictable'' in these systems, hence
not ``complex'', although the single trajectories may display arbitrarily high complexity (in symbolic dynamics,
and information theoretic terms, for instance).

Systems with an unpredictability which affects not only the single trajectories, for fixed geometry, 
or the transport coefficients, for uncertain geometry, but affects the transport laws themselves, would
seem to belong to a class with a higher level of ``complexity''. The polygonal channel studied in this 
paper is such a system. Other map systems with such dependency have also been observed --- see Ref.\cite{Kla05} 
(where the geometry unpredictably gave periodic, diffusive or ballistic behaviour) and Ref.\cite{Ott93}. 
Our numerical results indicate that the transport law can change from diffusive to sub-diffusive regimes, 
or alternatively quite high superdiffusive regimes, after small changes of the parameters defining the 
geometry. To quantify this kind of complexity, we propose the following definitions.

\vskip 5pt
\noindent
{\bf Definition 6. }{\it Consider a transport model, whose geometry is determined by the parameter
$y$, which ranges in the interval $[0,h]$, and such that its transport law is given by 
\be
\lim_{t \to \infty} \frac{\langle s_x^2(t) \rangle}{t^\zg} = A~, \quad 0 < A < \infty
\label{tr-law}
\ee
with $\zg$ a function of $y$ varying in $[0,2]$, when $y$ spans $[0,h]$. Let $\Delta \zg(y_m,y_M) \in [0,\infty]$ 
be the difference between the largest and the smallest value of $\zg$, for $y$ in the subinterval
$(y_m,y_M) \subset [0,h]$, where $\Delta \zg(y_m,y_M) = \infty$ if in $(y_m,y_M)$ there are points
for which \q{tr-law} is not satisfied by any $\zg \ge 0$.

\begin{itemize}

\item[{\bf i.}] The {\em transport complexity of first kind} of the transport model in $(y_m,y_M)$ is 
the number
\be
\mathcal{C}_1(y_m,y_M) = \frac{h \Delta \zg(y_m,y_M)}{2(y_M-y_m)} \in [0,\infty)
\label{C1}
\ee
if it exists.
\item[{\bf ii.}] The {\em transport complexity of second kind} of the transport model for
$y=\hat{y}$ is the exponent $\mathcal{C}_2=\mathcal{C}_2(\hat{y})$, if it exists, for which the limit
\be
\lim_{\varepsilon \to 0} 
\frac{\mathcal{C}_1(\hat{y}-\varepsilon,\hat{y}+\varepsilon)}{\varepsilon^{\mathcal{C}_2(\hat{y})}} 
\label{C2}
\ee
is finite.
\item[{\bf iii.}] The {\em transport complexity of third kind} of the transport model for
$y=\hat{y}$ is the limit 
\be
\mathcal{C}_3(\hat{y}) = \lim_{\varepsilon \to 0} \Delta \zg(\hat{y}-\varepsilon,\hat{y}+\varepsilon)
\ee
\end{itemize}
}

\vskip 5pt
\noi
These definitions are motivated by the following considerations. If $\mathcal{C}_1$ 
does not vanish, the system is surely highly unpredictable from the point of view
of transport, in the interval $(y_m,y_M)$, because its transport law is only known with 
a given uncertainty. This could be the case, for instance, of a batch of microporous
membranes with flat pore walls, whose orientation is obtained with a certain tolerance
(transport complexity of first kind). However, $\mathcal{C}_1$ may diverge
around some point of $[0,h]$, as our data seem to indicate, giving rise to an even higher
level of transport complexity. Assuming that this divergence has 
the form of a power law, we take the power as a measure of this second kind of complexity.
But even this level of complexity seems to be insufficient for our models,
which indicate that $\Delta \zg(y_m,y_M)$ could be discontinuous. For this
reason, we introduce the third kind of transport complexity.

In order to further investigate these notions, and how they relate to our systems, we have
considered the transport behaviour in a narrow range about two principal systems, taken from the
examples of super-diffusive transport observed in Section \ref{equilibrium}. The first principal
system is the rational parallel $\Delta y_t/\Delta x = 1$, where we consider the transport behaviour
in the limit that $\Delta y_b/\Delta x \rightarrow 1, \Delta y_b/\Delta x>1$. The second principal
is the irrational parallel $\Delta y_t/\Delta x = 2$ system, where we consider the transport
behaviour in the limit that $\Delta y_b/\Delta x \rightarrow 2, \Delta y_b/\Delta x>2$. In
\tabl{complex-tab}, we show the exponent $\gamma$ of the transport laws (as per \q{tr-law})
corresponding to the various choices for $\Delta y_t$ and $\Delta y_b$. 

\begin{table}
  \centering
  \begin{tabular}{c|c|c||c|c|c}
    $\Delta y_t/\Delta x$ & $\Delta y_b/\Delta x$ & $\gamma$   &     $\Delta y_t/\Delta x$ & $\Delta y_b/\Delta x$ & $\gamma$    \\
\hline
         1       &    1.01      &    0.71(4)    &         2       &    2.02      &    1.04(2)      \\  
         1       &    1.001     &    0.35(6)    &         2       &    2.002     &    1.01(2)      \\  
         1       &    1.0001    &    0.66(5)    &         2       &    2.0002    &    1.04(2)      \\  
         1       &    1.00001   &    0.58(3)    &         2       &    2.00002   &    1.02(2)      \\  
         1       &    1.000001  &    0.53(5)    &         2       &    2.000002  &    0.98(2)      \\
         1       &    1         &    1.66(3)    &         2       &    2         &    1.83(3)      
  \end{tabular}
  \caption{Transport law exponent $\gamma$, as identified by \q{tr-law}, observed for various
  choices of $\Delta y_t/\Delta x$ and $\Delta y_b/\Delta x$. There appears to be no trend toward super-diffusive
  transport, arbitrarily close to (but away from) the parallel case. Numbers in brackets indicate
  errors in the last figure.} 
  \label{complex-tab}
\end{table}

We note that, in both cases, there appears to be a strong discontinuity at $\Delta y_t \Delta y_b$ --- when the walls are not parallel, the behaviour is no longer superdiffusive. In the
rational case, the transport coefficient appears to depend unpredictably on the $\Delta y_t$, but is
always distinctly \emph{sub-diffusive}. By contrast, in the irrational case, the transport appears
to be essentially diffusive when the walls are not parallel. This behaviour is consistent with 
that observed in Section \ref{unpar-coll-eq} for the unparallel walls, which are also irrational. 

Due to the discontinuity, the transport complexities $\mathcal{C}_1$ and  $\mathcal{C}_2$ diverge in
these cases, reflecting the high degree of unpredictability that has been observed in the previous
sections, which can only be quantified by $\mathcal{C}_3$.


\section{Conclusion}
\label{conclude}

In this paper, we have examined the transport properties of a dynamical system of remarkable simplicity --- a 
two-dimensional channel with straight-edged walls, populated by non-interacting, point-like particles. 
Importantly, our choice of system ensures the absence of dynamical chaos. However, this system displays a rich 
variety of transport behaviours, belying the simplicity of the underlying dynamics.

For the parallel equilibrium systems where the pore angles $\theta_t$ and $\theta_b$ are equal ($\theta_t = 
\theta_b = \theta$), we observe super-diffusive (but not ballistic) behaviour. For the parallel
irrational polygonal billiards, we find that the root-mean-square displacement grows approximately as 
$\ave{s_x^2(t)}\sim t^{1.85}$ in the equilibrium dynamics. However, for the rational counterpart,
$\ave{s_x^2(t)}\sim t^{1.65}$ at equilibrium, apparently constrained by the limited range of
accessible momenta. These results hold also for the case where $\theta_t=0, \theta_b=\theta$, which
are isomorphic to systems where the top and bottom pore angles are equal, but the saw-teeth are
$\Delta x$ out of phase. Interestingly, these results are not strongly sensitive to the opening of
the horizon (particularly in the case of a shallow saw-tooth angle, $\theta<\pi/4$). It is perhaps
surprising that no ballistic motion was observed in these systems, given their simple structure ---
while the transport is strongly super-diffusive, it is also clearly sub-ballistic in nature.

The behaviour for individual trajectories in these parallel-walled systems are strongly dependent on
the initial conditions, in part because of the nature in which the dynamics permits (or rather,
restricts) the exploration of the momentum space. This can be understood through the strong correlation
of any particular value in the sequence of a particle's momenta with those momenta preceding (and
indeed, following) it, developing trajectory sequences through the creation of seemingly
quasi-periodic ``building blocks'' on various length scales. Despite the limitations imposed by this
strong correlation, the behaviour of the trajectory over a particular order of magnitude in time is
seemingly unpredictable from its behaviour over shorter time-scales, and it appears that the
correlation length is of the order of $10^9$ time units, or more, when $\theta$ is irrational with
respect to $\pi$, and orders of magnitude longer when $\theta$ is rationally related to $\pi$. To
give some physical sense to this observation, we consider an analogy to the transport of a light symmetric
gas molecule (such as methane) along a nanopore of width 1 nm. At room temperature, the unit
velocity in our system corresponds to a root-mean-square velocity for methane of approximately $400$
m/s, implying that our time units correspond roughly to units of picoseconds. Consequently, the
correlation length in the irrational systems is at least of the order of microseconds, during which time the
particle will have travelled of the order of millimetres. For the rational systems, the lower bound for
the correlation time is closer to the order of seconds, although the mean displacement over this time is 
of similar order (because of the slower transport, reflected by the smaller value of $\zg$). We note that 
the mean time between collisions would be much less than this, for all but the most rarified of conditions. 

For the nonequilibrium parallel systems, we observe similarly strong dependence on the initial conditions 
of the trajectories --- indeed, at some field strengths we observe clearly non-ergodic behaviour, with the 
existence of distinct attracting periodic trajectories (with different net speeds) for a single dynamical system. It 
appears that the existence of periodic orbits is determined by the field strength, although bursts of 
quasi-periodic behaviour are observed over short time intervals. There is no clear relation between the 
onset time for ballistic (periodic) transport and the field strength. More fundamentally, there does not 
appear to be any clear connection between the non-zero-field estimates of the diffusion coefficient and the 
zero-field (equilibrium) value (which is infinite in the case of parallel walls). Certainly, there is no 
evident divergence of $D(\epsilon)$ in the zero-field limit, as one would expect for a super-diffusive 
thermodynamic system (such as plug flow). Although, these systems could have a diffusion coefficient 
defined as in section IV, the existence of a linear regime has not been clearly estabilished, because the 
equality between $D$ and $D_{\text{ne}}$ is not guaranteed by our calculations.

For the unparallel systems, where $\theta_t\not=\theta_b$, we observed diffusive transport in the
 equilibrium system --- $\ave{s_x^2(t)}\sim t$, with a Gaussian distribution of displacements $s_x$ 
 consistent with the behaviour predicted by \q{gaussian}. The trajectories were more sensitive to the
 opening of the horizon than for the parallel systems, with occasional short-lived bursts of much
 faster transport. In contrast to the unexpected richness in behaviour of the parallel systems, in the
 unparallel systems, the
 properties of individual trajectories consistently matched ensemble properties. This was also true
 of the nonequilibrium results, where there was a much clearer field-dependence of the transport
 behaviour, and in particular the transition to a periodic orbit. In the unparallel systems, the
 periodic orbit was unique for a given field (and appeared to be continuously dependent on the
 field), and the mean onset time of the periodic orbit grew with decreasing field strength. This is
 consistent with the results of Ref. \cite{jsp099_0857}, which, however, were obtained for a system
 with \emph{parallel} boundary walls. However, the wall angles were chosen such that $\tan\theta$
 was close to the golden ratio $(\sqrt{5}+1)/2=1.618\ldots$, so as to ``maximise'' the irrationality
 of the relationship of $\theta$ to $\pi$. It is possible that such a choice leads to much smaller
 correlation times in the momentum sequence for particle trajectories, and consequently faster
 convergence to the longer time-scale behaviour more reminiscent of intermolecular collisions,
 observed in \fig{par-s2}. 

Despite the apparently ergodic behaviour, however, our results are not conclusive regarding the
connection between the equilibrium and nonequilibrium transport behaviours: the existence of a
linear response regime, where the non-zero-field estimates $D(\epsilon)$ converge to the equilibrium
diffusion coefficient in the zero-field limit is not clear.

Among the systems we have observed, we find that: 
\begin{itemize}
\item systems with irrational walls are super-diffusive 
if parallel, but diffusive if rationally related; 
\item systems with one rational wall and one irrational
wall are sub-diffusive; 
\item systems with two rational walls are super-diffusive. 
\end{itemize}

Furthermore, we identify a means of quantitatively estimating the uncertainty related to the system
transport, through the parameters used to define the geometry. Such notions of transport complexity
are useful in distinguishing between various transport systems, the range of transport properties
that they exhibit, and in particular the predictability of these transport properties. The various
transport properties observed in the systems examined in this paper, and their strongly
unpredictable nature, are reflected in the divergent values for the proposed transport complexities
$\mathcal{C}_1$ and $\mathcal{C}_2$, and in the finite value of $\mathcal{C}_3$. 
We note that, for the sake of simplicity, we have focussed on the dependence of the
transport law on just one paramenter, but our investigation reveals that
transport in our models may depend in a counterintuitive and irregular
fashion on other parameters as well, such as $h$. Striking is the case in which
a growth of $h$ implies a reduction of the transport exponent. More complicated
notions of transport complexity may then be envisaged, but the conceptual picture 
outlined here would not change.

This analysis leads us to the following observation. As is well known, the thermodynamical 
properties of a macroscopic system are not a function of the boundary of the medium; the 
properties of a fluid, such as its viscosity, do not depend on the
geometry of its container, and,  in general, the nature of transport is essentially independent of
the geometry. Fermi expressed this concept, at the beginning of his book on
thermodynamics \cite{Fermi}, in which he states: {\em ``The geometry of our system is obviously
characterized not only by its volume,  but also its shape. However, most thermodynamical
properties are largely independent of the shape, and, therefore, the volume is the only
geometrical datum that is ordinarily given. It is only in the cases for which the ratio of surface
to volume is very large (for example, a finely grained substance) that the surface must be
considered.} This is clearly understood in the terms of kinetic theory, which leads to
explicit expressions for the mutual or self diffusion coefficients in terms of only the mean-free paths
$\zl$ for collisions among particles, without any reference to the shape of the containers. Only
in the  case that $\zl$ is of the same order of the characteristic lengths of the container, does
this play a role; but in that case, the standard laws of thermodynamics cease to hold, and are
replaced by those of highly rarefied gases, or Knudsen gases \cite{DGM,LR84,KK80}.
Because our systems do not have this extreme ratio of surface to
volume, our results indicate that our systems cannot be considered as thermodynamic systems, much as
they remain highly interesting from both theoretical and technological standpoints. Indeed, as has been noted
elsewhere \cite{CR02,RC02}, much care must be taken in the thermodynamic interpretation of the collective 
behaviour of systems of non-interacting particles. The arguments of Refs.\cite{CR02,RC02} mainly
referred to nonequilibrium systems: here we provide some examples which show that they apply to
equilibrium systems as well. All this can be summarized stating that a
certain degree of {\em chaos}, or of {\em randomness} is required in the
microscopic dynamics of particle systems, for them to look like thermodynamic systems,
but there are two ways in which this can be achieved. If the randomness is
intrinsic to the fluid, the geometry of the system is a secondary issue, and
the fluid behaves as a proper thermodynamic system. If the randomness is produced by
the geometry of the outer environment, the geometry plays an obviously
important role, and systems such as ours do not behave like proper thermodynamic
systems.


\section{Acknowledgements}
\label{ack}

We would like to thank R. Artuso, M. Falcioni, R. Klages and A. Vulpiani for helpful 
feedback from preliminary drafts. We would also like to thank the ISI Foundation
for financial support throughout this work.



\end{document}